\documentclass[amsmath,amssymb,aps,prl,superscriptaddress,twocolumn,floatfix]{revtex4}
\usepackage{graphicx}
\usepackage{dcolumn}
\usepackage{bm}
\usepackage[usenames]{color}
\usepackage{comment}

\usepackage{hyperref} 
\usepackage{physics}
\usepackage{bookmark}
\usepackage{doubleangle}
\usepackage{mathtools}
\usepackage{subfigure}

\begin{document}

\title{Hamiltonian reconstruction as metric for  variational studies}
\author{Kevin Zhang}
\affiliation{Laboratory of Atomic and Solid State Physics, Cornell University, 142 Sciences Drive, Ithaca NY 14853-2501, USA}
\author{Samuel Lederer}
\affiliation{Laboratory of Atomic and Solid State Physics, Cornell University, 142 Sciences Drive, Ithaca NY 14853-2501, USA}
\author{Kenny Choo}
\affiliation{Department of Physics, University of Zurich, Winterthurerstrasse 190, 8057 Zurich, Switzerland}
\author{Titus Neupert}
\affiliation{Department of Physics, University of Zurich, Winterthurerstrasse 190, 8057 Zurich, Switzerland}
\author{Giuseppe Carleo}
\affiliation{Institute of Physics, \'{E}cole Polytechnique F\'{e}d\'{e}rale de Lausanne (EPFL), CH-1015 Lausanne, Switzerland}
\author{Eun-Ah Kim}
\affiliation{Laboratory of Atomic and Solid State Physics, Cornell University, 142 Sciences Drive, Ithaca NY 14853-2501, USA}

\date{January 2021}

\begin{abstract}
Variational approaches are among the most powerful modern techniques to approximately solve quantum many-body problems. These encompass both  variational states based on tensor or neural networks, and parameterized quantum circuits in variational quantum eigensolvers. However, self-consistent evaluation of the quality of variational wavefunctions is a notoriously hard task. 
 Using a recently developed Hamiltonian reconstruction method, we propose a multi-faceted approach to evaluating the quality of neural-network based wavefunctions. 
 Specifically, we consider convolutional neural network (CNN) and restricted Boltzmann machine (RBM) states trained on a square lattice spin-1/2 $J_1$-$J_2$ Heisenberg model.
 We find that the reconstructed Hamiltonians are typically less frustrated, and have easy-axis anisotropy near the high frustration point. Furthermore, the reconstructed Hamiltonians suppress quantum fluctuations in the large $J_2$ limit. Our results highlight the critical importance of the wavefunction's symmetry. Moreover,
 the multi-faceted insight from the Hamiltonian reconstruction reveals that a variational wave function can fail to capture the true ground state through suppression of quantum fluctuations. 

\end{abstract}

\maketitle

\begin{figure}[t]
    \centering
    \includegraphics[width=8.5cm]{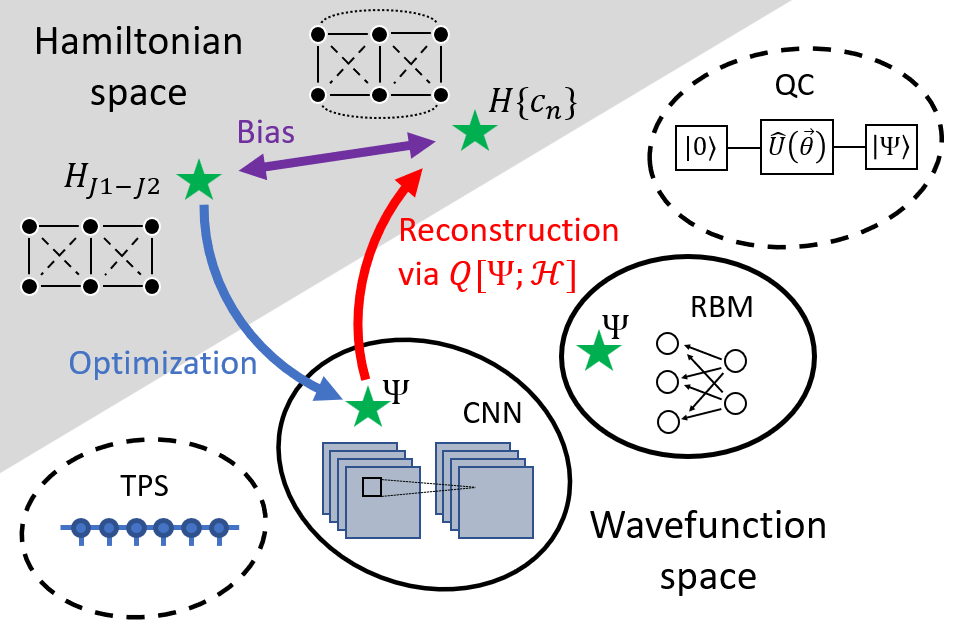}
    \caption{In a typical variational algorithm, a wavefunction is obtained through variational optimization within a given variational form such as CNN, RBM, tensor product state (TPS), or a parametrized quantum circuit (QC).
    In this work, we study CNN and RBM quantum states, marked with green stars.
    A blue arrow is shown to represent variational optimization of the CNN construction as an example.
    The Hamiltonian reconstruction works in the opposite direction to map a variational wavefunction 
    to a Hamiltonian $H[\{c_n\}]$ (red arrow).
    The bias between the original Hamiltonian and the reconstructed Hamiltonian (purple arrow) provides insight into the nature of the variational wavefunction.}
    \label{fig:overview}
\end{figure}

{\it Introduction --} The Hamiltonian is the defining object
that governs the dynamics of a physical system. For a quantum mechanical system, 
it defines the Schr\"odinger equation to be solved to obtain the energy spectrum and the wavefunction. However, the approach of ``exact diagonalization" is constrained to small system sizes due to the exponential growth of the Hilbert space upon increasing the system size.
An alternative to exact diagonalization is the Quantum Monte Carlo techniques using a stochastic approach to model the probability distribution associated with the thermal density matrix associated with a given Hamiltonian. These approaches, however, suffer from the sign-problem \cite{PhysRevLett.94.170201}, which limits their applicability to a restricted class of Hamiltonians, or to high temperature properties only. 
These challenges motivated variational wavefunction approaches to start from many-body wave functions that are parameterized within a given functional form.
In variational approaches, the Hamiltonian is referenced for optimizing the wavefunction within the chosen functional form (see the blue arrow in \autoref{fig:overview}). Since the resulting best wavefunction is constrained to lie within limited variational spaces such as 
tensor network states \cite{doi:10.1080/14789940801912366}, neural network states \cite{Carleo602,PhysRevB.100.125124}, and parametrized quantum circuits \cite{Peruzzo_2014,McClean_2016} (see~\autoref{fig:overview}), significant effort has been put into having sufficiently general variational classes that can capture the actual ground state. However, assessing how close a given variational parameterization is to the target ground state is, in general, a hard task.

At present, the standard metrics for assessing the quality of a wavefunction that cut across different variational forms are the energy and the energy variance. Reliance on these measurements, however, leaves the comparison between constructions a case-by-case trial exercise.
Much needed are alternative metrics to assess the quality of a given variational state.
Interestingly, recent works have proposed methods to reconstruct Hamiltonians from measurements of correlators \cite{Qi2019determininglocal,PhysRevX.8.031029,PhysRevLett.122.020504,Valenti_2019} or single operator measurements \cite{Pakrouski2020automaticdesignof} (see the red arrow in  \autoref{fig:overview}).
These reconstruction processes have been tested on Hamiltonians with known exact solutions, but their applicability to challenging open problems has yet to be demonstrated.

In this Letter, we employ Hamiltonian reconstruction to investigate how frustration affects the distance between reconstructed and target Hamiltonians for neural-network wavefunctions. 
We optimize convolutional neural network (CNN) and restricted Boltzmann machine (RBM) wavefunctions to approximate the ground state of 
the spin-1/2 $J_1$-$J_2$ Heisenberg model on a square lattice~\cite{PhysRevB.100.125124}, a poster-child frustrated spin model that suffers from the sign problem.
 On this model, deep neural network-based wavefunctions have so far obtained highly accurate results for the $J_1$-$J_2$ model away from the high frustration point, showing the potential of neural-network based variational constructions.
However, the same neural network states showed limitations near $J_2/J_1 = 0.5$, which is the point of high frustration \cite{PhysRevB.100.125124}. 
To probe features of these wavefunctions, we construct subspaces of Hamiltonians that accommodate different ``deformations" of the target Hamiltonian.
For each subspace, we use the reconstruction method to retrieve the Hamiltonian that best fits the trained wavefunction.
We then discuss insights from the Hamiltonian reconstruction. 

{\it Hamiltonian Reconstruction --} Let us review the  Hamiltonian reconstruction procedure following Refs.~\cite{Qi2019determininglocal,PhysRevX.8.031029}. The procedure starts with the wavefunction of interest $\Psi$, which would be energy-optimized within a given variational form.
We then define the Hamiltonian subspace, to be searched within, by a spanning set of operators $\mathcal{O}\equiv\{O_i\}$. Any Hamiltonian that is an element of this subspace, i.e., $H\in\mathcal{H}$, can be expressed in the form
\begin{equation}
    \label{eq:ham-repr}
    H[\{c_n\}] = \sum_n^N c_n O_n,
\end{equation}
where $c_n$'s are real-valued parameters.
The aim of the reconstruction procedure is to find the $N$-dimensional vector $\{c_n\}$ such that the wavefunction of interest $|\Psi\rangle$ is an approximate eigenstate of the corresponding Hamiltonian $H[\{c_n\}]$. For this, we construct the quantum covariance matrix $Q$ associated with the wavefunction and the Hamiltonian subspace 
\begin{equation}
\label{eq:qcm}
    Q[\Psi;\mathcal{H}]_{nm} = \frac{1}{2} \left( \langle O_n O_m \rangle +  \langle O_m O_n \rangle \right ) - \langle O_n\rangle \langle O_m \rangle,
\end{equation}
which is an $N\cross N$ positive semi-definite matrix where expectation values are evaluated with respect to the  wavefunction $|\Psi\rangle$ (also see Figure~\ref{fig:overview}).
The computational cost of evaluating $Q$ is quadratic in the number of operators to be considered (but note that the number of terms in operators also tends to grow linearly with system size). 

The subset of Hamiltonians $\mathcal{H}[\mathcal{O}]$ that correspond to eigenvectors of $Q[\Psi;\mathcal{H}]$ with the smallest eigenvalues would all accept $|\Psi\rangle$ as an approximate eigenstate.
To see this, note that the variance of the Hamiltonian $H[\{c_n\}]$ in the state $\ket{\Psi}$ is given by
\begin{equation}
\begin{split}
    \langle(\Delta H[\{c_n\}])^2\ &= \langle H[\{c_n\}]^2 \rangle - \langle H[\{c_n\}] \rangle^2 \\
    &= \sum_{nm} c_n c_m \left(\langle O_n O_m \rangle - \langle O_n \rangle \langle O_m \rangle \right) \\
    &= \vec{c}^{~T} Q[\Psi; \mathcal{H}] \vec{c}.
\end{split}
\end{equation}
By diagonalizing $Q[\Psi;\mathcal{H}]$, the Hamiltonians $H[\{c_n\}]$ which have the lowest variance under $\ket{\Psi}$ can be found, and the associated eigenvalues will be the variances in energy of those Hamiltonians.
If $|\Psi\rangle$ is an exact ground state of the exact parent Hamiltonian $H^*$, and $H^*$ is within the Hamiltonian search space $\mathcal{H}[\mathcal{O}]$, then $H^*$ will lie in the nullspace of $Q[\Psi;\mathcal{H}]$. 

The expectation values of many-body operators in Eq.~\eqref{eq:qcm} need to be evaluated by performing high-dimensional integrals.
Typically, these high-dimensional integrals can be approximated via Monte Carlo (MC) sampling, but we found that the Hamiltonian reconstruction is highly sensitive to noise in the correlation functions (see Supplemental Material II.C).
This sensitivity restricts the procedure to systems in which the correlation functions can be evaluated sufficiently accurately; indeed, previous applications of Hamiltonian reconstruction \cite{PhysRevX.8.031029,PhysRevB.99.020202} are limited to well-understood states in which correlation functions can be evaluated easily.
In our case, this restricted our study to small system sizes in which the correlation functions could be evaluated explicitly.

The antiferromagnetic $J_1$-$J_2$ model for spin 1/2 on a two-dimensional square lattice~\cite{PhysRevLett.63.2148,Schulz_1992} is defined by the  following Hamiltonian
\begin{equation}
   H_{J_1J_2} \equiv J_1 \sum_{\langle i j \rangle} \vec{S}_i \cdot \vec{S}_j + J_2 \sum_{\llangle i j \rrangle} \vec{S}_i \cdot \vec{S}_j,
   \label{eq:J1J2}
\end{equation}
where $\langle i j \rangle$ and $\llangle i j \rrangle$ denote nearest and next-nearest neighbours respectively.
We set $J_1=1$ and consider only antiferromagnetic interactions $J_2 \geq 0$. The exact ground states of the Hamiltonian in the two limits of $J_2\ll J_1$ and $J_2\gg J_1$ are well understood, since geometric frustration is absent in both limits: 
the ground state is a N\'{e}el antiferromagnet for $J_2\ll J_1$ and a striped antiferromagnet for $J_2\gg J_1$. 
 However, the nature of the ground state in the vicinity of the maximally frustrated point of $J_2/J_1=0.5$ is the subject of much debate \cite{WJHu2013, SSGong2014, HCJiang2012, Sachdev1990, Mambrini2006,JFYu2012, LWang2018, FFerrari2020, YNomura2020}.

{\it Hamiltonian space and wavefunction space --} We consider three Hamiltonian subspaces that allow the reconstructed Hamiltonian to deviate from the target Hamiltonian Eq.~\eqref{eq:J1J2} in physically meaningful ways.
We chose the three two-operator parametrizations
\begin{equation}
\label{eq:spaces}
    \begin{split}
        H[\delta J_2] &= H_{J_1J_2} + \delta J_2 \sum_{\llangle i, j\rrangle} \vec{S}_i \cdot \vec{S}_j \\
        H[J_3] &= H_{J_1J_2} + J_3 \sum_{\left< i, j\right>_3} \vec{S}_i \cdot \vec{S}_j \\
        H[\alpha] &= H_{J_1J_2} + \alpha \left( \sum_{\left< i, j\right>} S^z_i S^z_{j} + \frac{J_2}{J_1} \sum_{\llangle i, j\rrangle} S^z_i S^z_{j} \right),\\
    \end{split}
\end{equation}
where $\delta J_2$ measures the degree of frustration, $J_3$ measures the strength of a longer range interaction, and $\alpha$ measures the easy-axis anisotropy.
Above, the coefficients of the original $J_1$-$J_2$ Hamiltonian have been normalized to 1.
In Supplemental Material -- II.A, we show that reconstructions into these spaces, from an exact solution of the Heisenberg model, simply yield the original Heisenberg model.
In Supplemental Material -- II.B, we present how reconstructions into higher-dimensional spaces are more challenging to interpret, motivating our choice of study of two-dimensional Hamiltonian spaces.

Seeking further understanding of the challenges underlying the maximally frustrated point, we focus on neural network based wavefunctions that outperformed (i.e., had lower energy than) leading variational constructions, away from the high frustration point~\cite{PhysRevB.100.125124}. 
Neural networks can be universal approximators of complex functions \cite{cybenko1989,bishop-book} and thus have the potential to allow more efficient exploration of the wavefunction space compared to traditional constructions \cite{deng2017prx}. 
The initial proposal of using restricted Boltzmann machines (RBM) to represent many-body wavefunctions \cite{Carleo602} generated much excitement in the community and extensive investigations of RBM-based wavefunctions and their variants \cite{PhysRevB.96.205152,carleo_nomura_imada_2018,PhysRevB.97.085104,PhysRevX.8.011006,doi:10.7566/JPSJ.87.014001,PhysRevLett.121.167204,kochkov2018variational,Luo_2019,PhysRevB.99.165123,sharir2020deep,nomura2020diractype,Gao_2017}.  More recently, \textcite{yoav2019prl} showed that the more expressive convolutional neural network (CNN) architecture can encode volume-law entangled states more efficiently. Indeed, CNN wavefunctions improved energy compared to state-of-the-art methods for the $J_1$-$J_2$ model, but only in the parameter regime away from the classical frustration point of $J_2/J_1=0.5$ \cite{PhysRevB.100.125124}.

In this work, we examine CNN and RBM many-body wavefunctions which were trained to to have high overlap with the ground state of the model in question, the spin-1/2 Heisenberg $J_1$-$J_2$ model on a $4\times 4$ square lattice with periodic boundary conditions (see Supplemental Material I).
Both the CNN and RBM architectures preserve the translational invariance of the system, and the wavefunctions were further symmetrized to respect time reversal and point group symmetries (see Supplemental Material -- I.C).
We trained wavefunctions for values of $J_2$ ranging between 0 and 2. The optimization of the wavefunctions was done using the NetKet package \cite{netket}.
 
\begin{figure*}[htp]
   \subfigure[] {
    \centering
    \includegraphics[width=.4\textwidth]{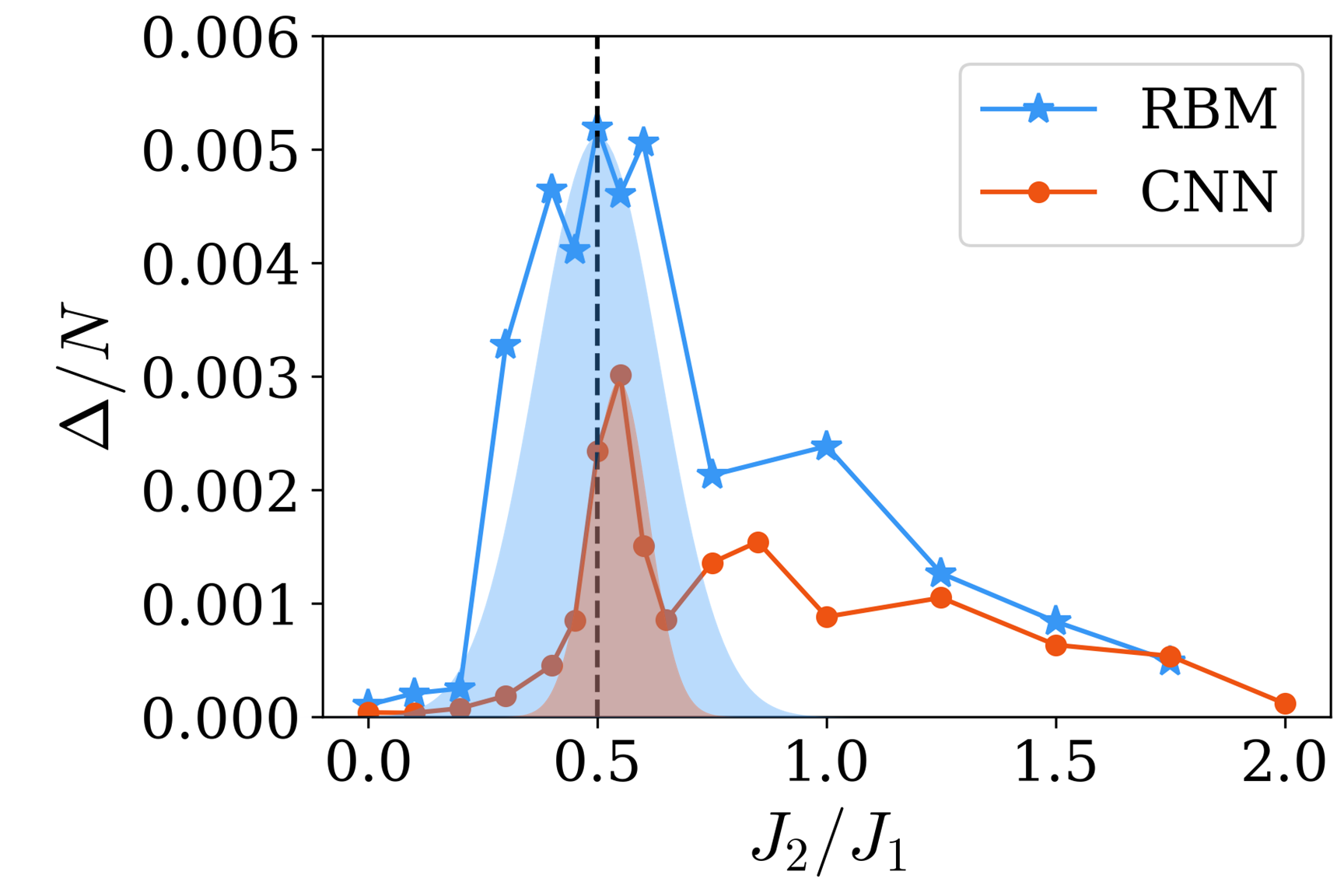}
    }\subfigure[] {
    \centering
    \includegraphics[width=.4\textwidth]{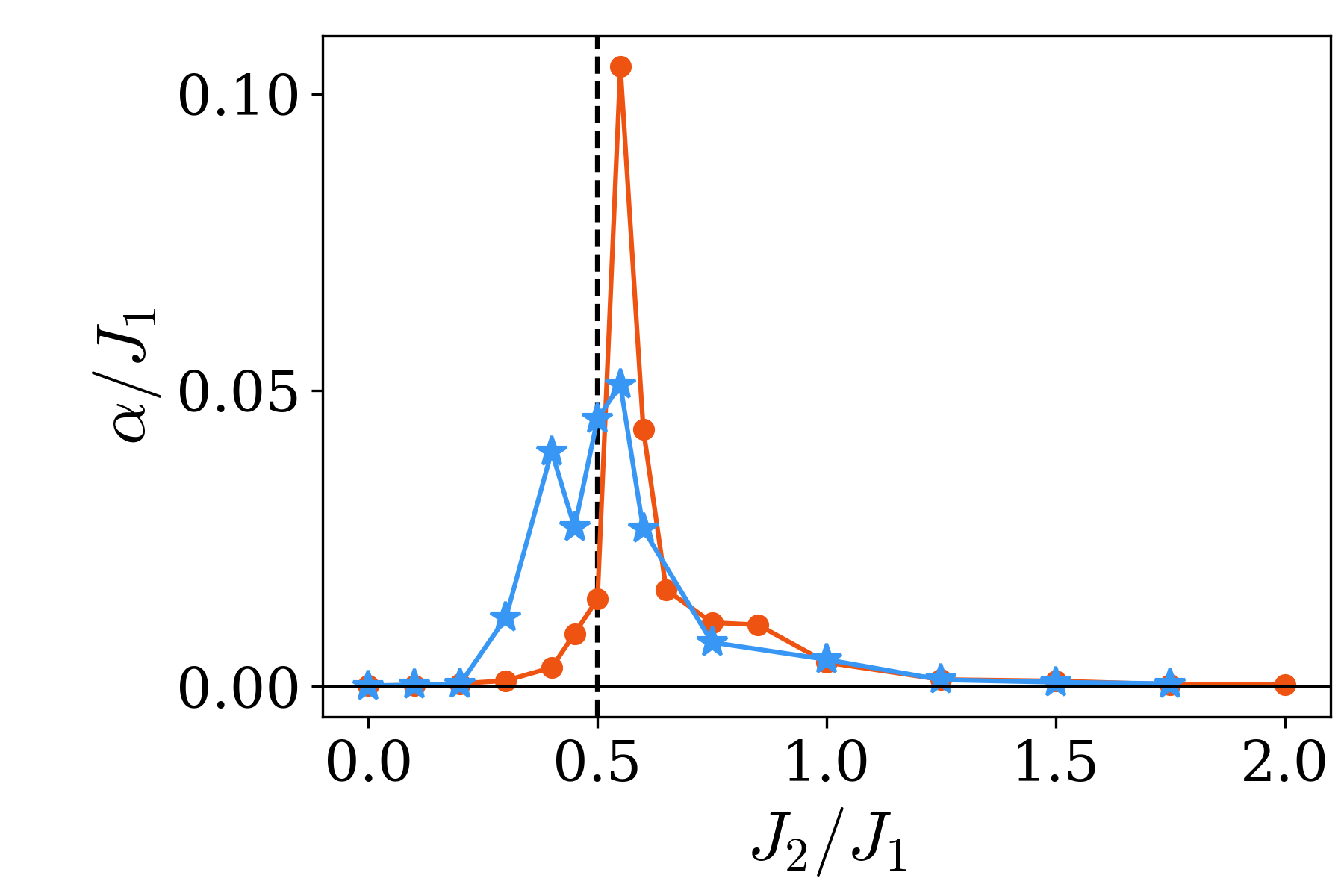}
    }
    \subfigure[] {
    \centering
    \includegraphics[width=.4\textwidth]{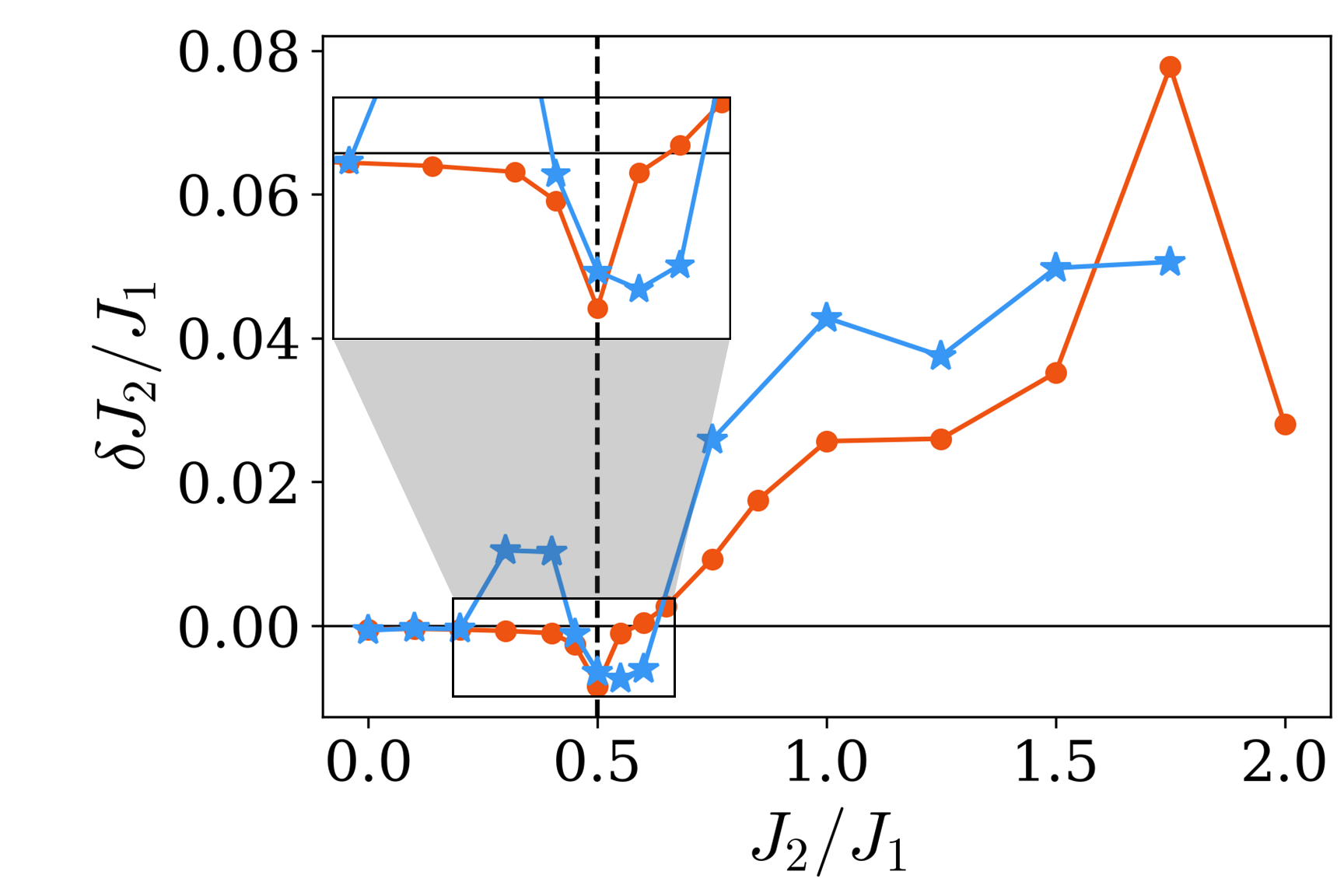}
    }\subfigure[] {
    \centering
    \includegraphics[width=.4\textwidth]{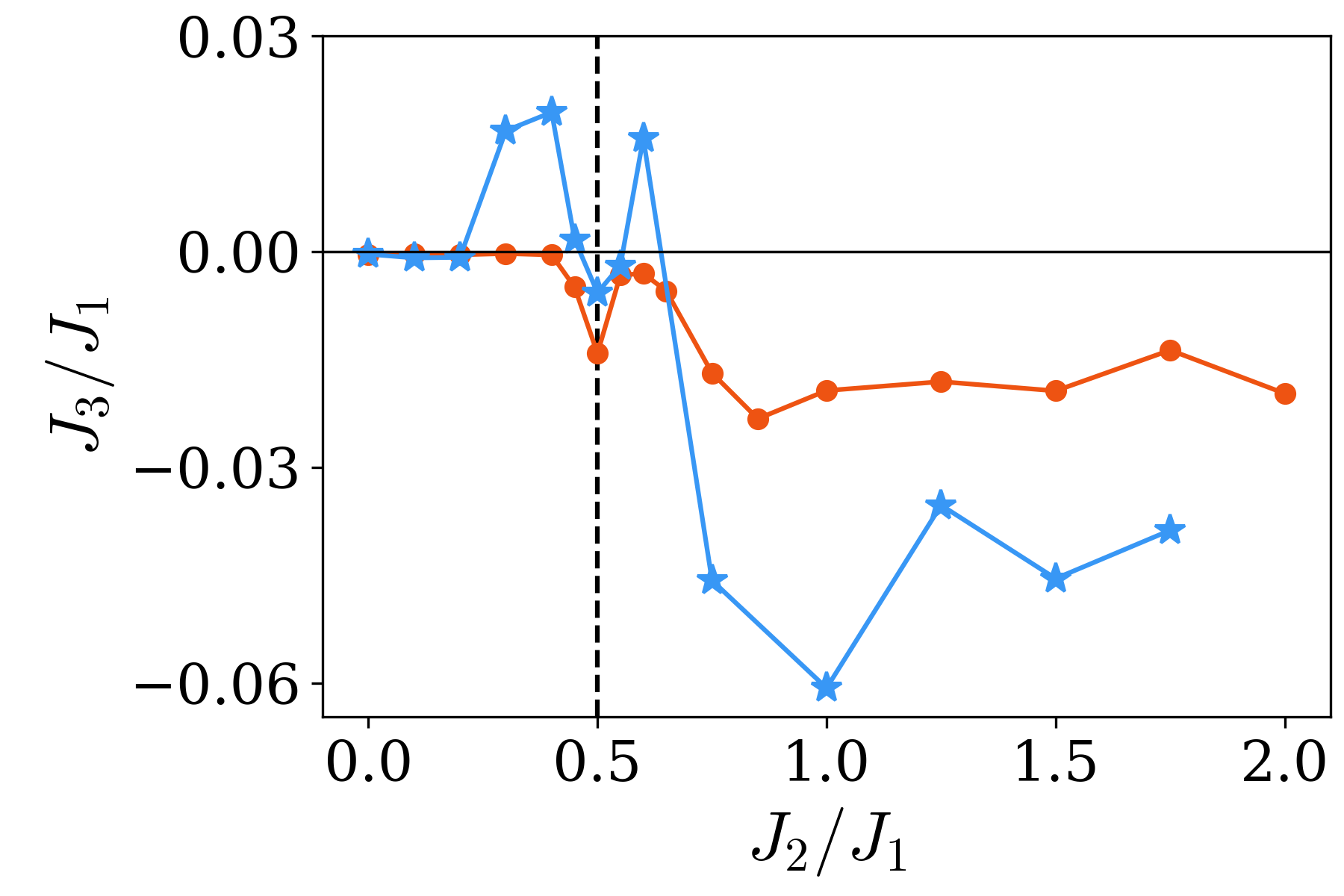}
    }
    \caption{Ground state energy differences, and parameters of the reconstructed Hamiltonians, of CNN and RBM wavefunctions.
    The vertical broken line marks $J_2/J_1 = 0.5$, which is the classical maximum frustration point.
    a) Ground state energy difference (relative to true ground state obtained via exact diagonalization) per site.
    The shaded area is a guide to the eye that outlines the range in which the energy difference is attributable to the wavefunction anisotropy (see panel b).
    The unshaded energy difference in the large $J_2$ regime is associated with errors in the reconstructed spin couplings, panels c) and d).
    b) The reconstructed easy-axis anisotropy $\alpha/J_1$, which is primarily peaked at the classical high frustration point and dies off in the small $J_2$ and large $J_2$ limits.
    c) The reconstructed difference in the nearest-neighbor coupling, $\delta J_2/J_1$. The reconstruction deviates slightly from the expected value of 0 around $J_2 = 0.5$, as well as in the large $J_2$ regime in a more pronounced manner. 
    d) The reconstructed longer-range interaction parameter $J_3/J_1$. Together with panel c), the reconstructions of the spin couplings $J_2$ and $J_3$ are associated with the energy differences of the trained wavefunctions outside the neighborhood of the classical high frustration point.}
    \label{fig:H1}
\end{figure*}

{\it Results --} The conventional measure for a wavefunction's quality is its variational energy.
The energies of our trained wavefunctions, measured with respect to the exact ground states are shown in \autoref{fig:H1}(a); the high frustration region around $J_2 = 0.5$ is marked by a sharp increase in energy difference.
However, in the large $J_2$ regime, past the high frustration region, the energy difference remains large.
The non-trivial dependence of the energies on the $J_2/J_1$ ratio implies multiple tendencies at play, yet the total energy ``bundles" any and all possible issues into a single number. We therefore compare reconstruction results shown in  \mbox{\autoref{fig:H1}(b-d)} to the variational energy to gain much needed insight. 

The easy-axis anisotropy, $\alpha$,
is sharply peaked in the vicinity of the high frustration point $J_2/J_1 = 0.5$ (see \autoref{fig:H1}(b)). The comparison between the reconstructed anisotropy and the energy difference (\autoref{fig:H1}(a)) reveals that the higher energies of both the CNN and RBM wavefunctions in the narrow region in the vicinity of the high frustration point can be attributed to wavefunction anisotropy.
Such an observation reinforces the importance of building spin rotation symmetry into the wavefunction, as was indicated by the performance of $SU(2)$ symmetric RBM wavefunctions for the
ground state of the Heisenberg model~\cite{PhysRevLett.124.097201}.
It is interesting to note that the CNN wavefunctions have more significant anisotropy despite having better energies compared to the RBM wavefunctions, once again confirming that energy by itself is an insufficient measure for validating variational wavefunctions.
However, the anisotropy alone does not explain the higher energies in the large $J_2$ region. 

The reconstructions of interaction strengths $\delta J_2$ and $J_3$ present complementary information. They show deviations from the target Hamiltonian in two separate regions: the vicinity of the high frustration point  $J_2/J_1 = 0.5$, and the large $J_2$ region (see \autoref{fig:H1}(c-d)).
In the vicinity of the high frustration point, the reconstructions tend to avoid the  high frustration point of $J_2/J_1 = 0.5$. In the $\delta J_2$ space, this is evidenced by negative $\delta J_2$ near the high frustration point.
In the large $J_2$ region, both $\delta J_2$ and $J_3$ are reconstructed in a way to strengthen the stripe order and reduce quantum fluctuations.
Specifically, large and positive $\delta J_2$, and large and negative $J_3$, both favor classical stripe order as we show explicitly through exact diagonalization in Supplemental Material -- III.

\begin{figure*}[htp]
   \subfigure[] {
    \centering
    \includegraphics[width=.4\textwidth]{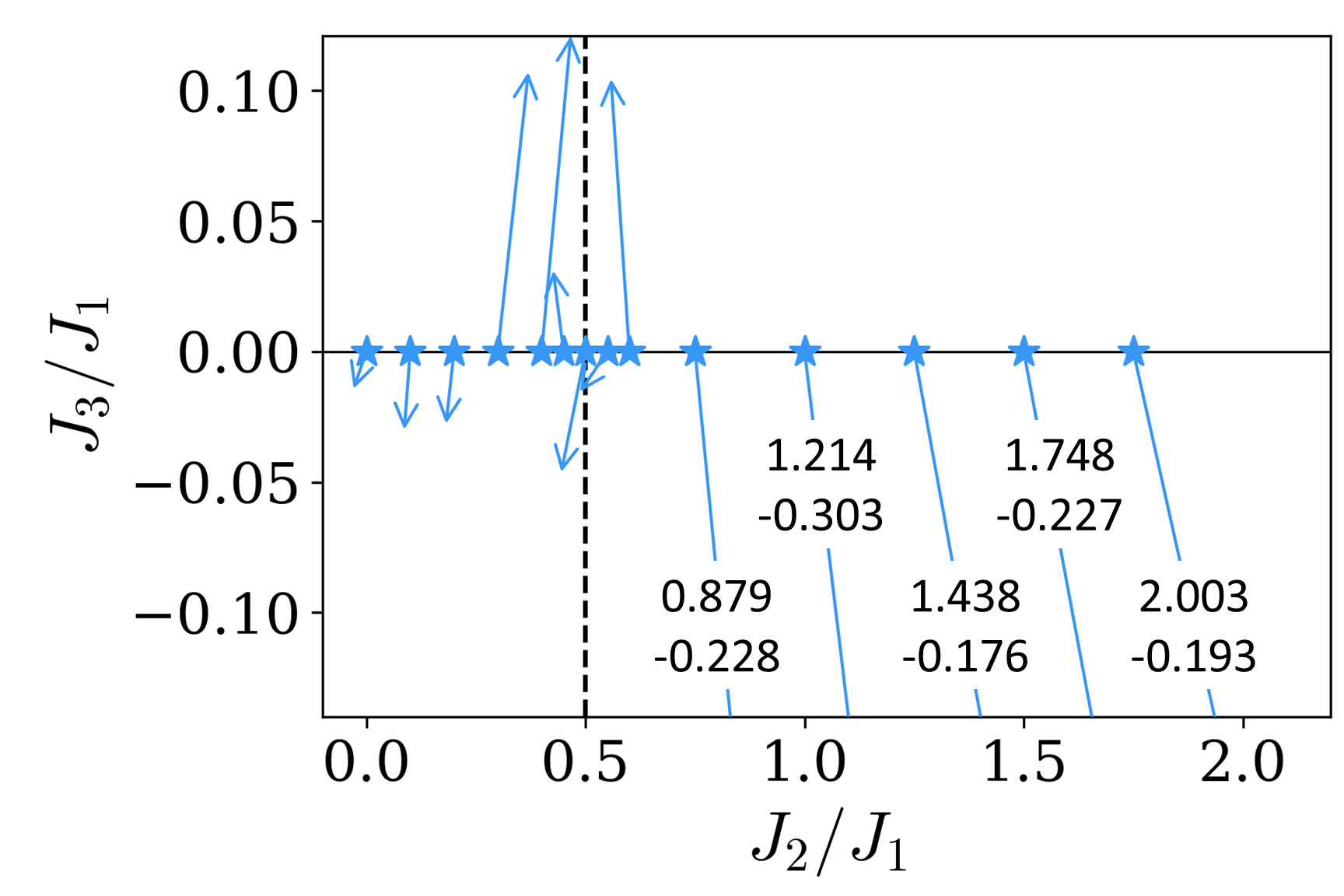}
    }
    \subfigure[] {
    \centering
    \includegraphics[width=.4\textwidth]{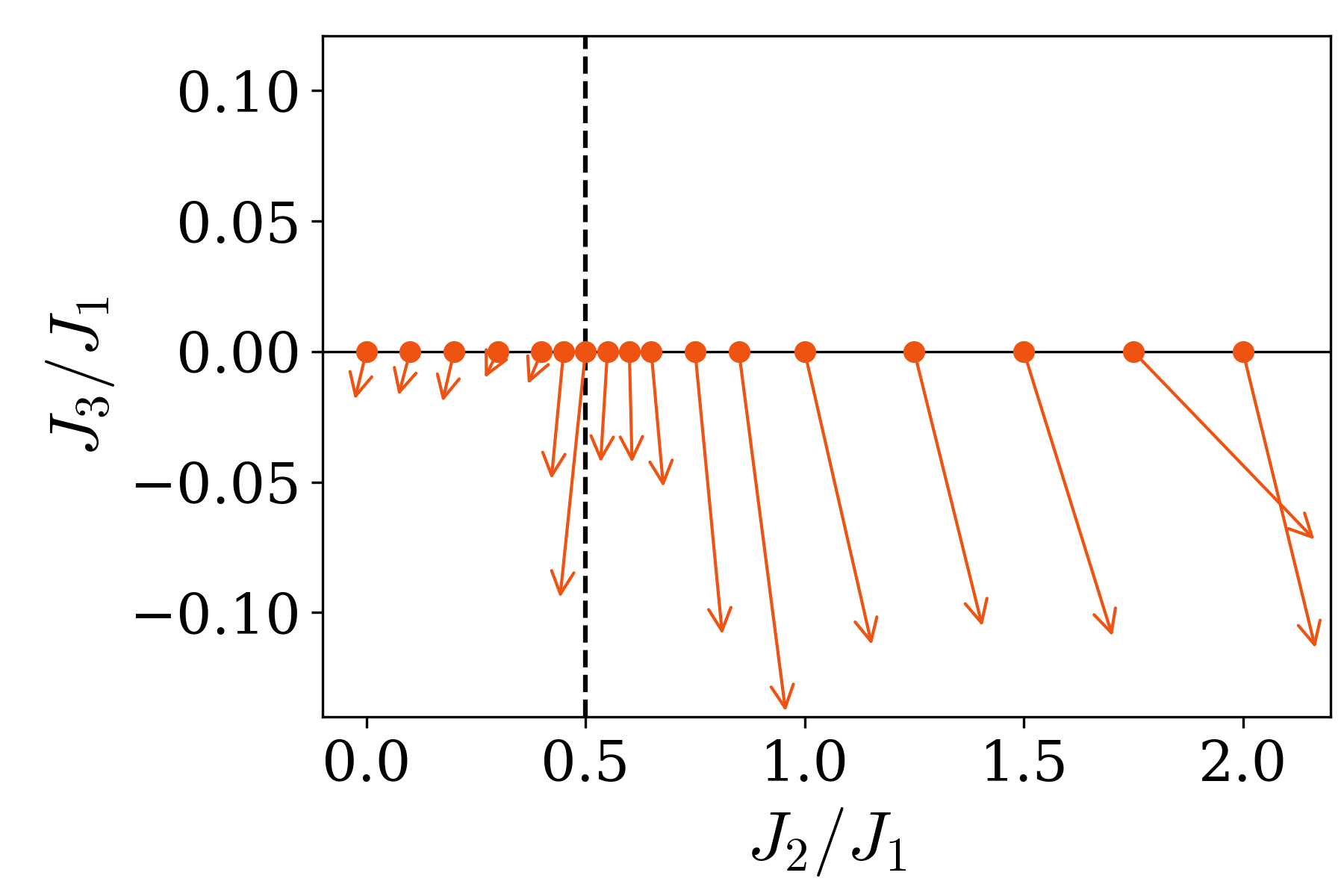}
    }
    \caption{Schematic summary of $\delta J_2$ and $J_3$ reconstructions of (a) RBM and (b) CNN wavefunctions.
    The markers represent initial $J_2/J_1$ parameters for which we trained variational wavefunctions.
    The tips of the arrows show the \textit{reconstructed} spin coupling parameters from the trained wavefunctions, i.e., $(\delta J_2 + J_2)/J_1$ and $J_3/J_1$, with the deviations magnified by a factor of 5 for clarity.
    The annotations beside clipped arrows describe the locations of the arrowheads (upper: $J_2/J_1$, lower: reconstructed $J_3/J_1$).
    }
    \label{fig:rg}
\end{figure*}

The compilation of the reconstruction results for these two parameters into one plot, as shown in \autoref{fig:rg}, demonstrates the tendency to avoid, or ``push away from" the high frustration point, as well as suppress quantum fluctuations via ferromagnetic $J_3$ at large $J_2$.
In other words, the reconstructions in the large $J_2$ regime that explain the large energy differences in Figure~\ref{fig:H1}(a) can be summarized as a general tendency to suppress quantum fluctuations.

{\it Conclusions --} We have proposed Hamiltonian reconstruction as a method to probe many-body variational wavefunctions beyond their energies.
Taking on the $J_1-J_2$ model and two neural network based variational wavefunctions, RBM and CNN, we investigated the Hamiltonian spaces parametrized by three channels of deviations from the target model: $\delta J_2$, $J_3$, and $\alpha$. 
Our results dissect the $J_2/J_1$ parameter space into two regimes: the regime dominated by frustration ($J_2/J_1 \approx 0.5$) and the regime dominated by classical stripe order ($J_2/J_1 > 0.5$). We found the anisotropy $\alpha$ to be the dominant cause of error near the high-frustration point. Moreover, we found $\delta J_2$ and $J_3$ reconstruction to both indicate suppression of quantum fluctuation through artificial enhancement of classical order in the large $J_2$ regime. Overall, the Hamiltonian reconstruction revealed mutliple ways for a variational wavefunction to fail in capturing highly frustrated ground states steeped in quantum fluctuations.

Looking ahead, we expect that Hamiltonian reconstruction can be an effective means to refine variational constructions in both classical and quantum (such as  variational quantum eigensolvers) platforms.
With new insight into the performance of variational wavefunctions, specific areas of improvement can be identified, informing future selection of variational constructions.
Further, our results concerning the $J_1$-$J_2$ model may serve as guidelines for future neural network studies of similar frustrated spin systems.

\noindent
{\bf Acknowledgements.}
KZ, SL, and E-AK acknowledge NSF, Institutes for Data-Intensive Research in Science and Engineering – Frameworks (OAC-19347141934714).
KC and TN acknowledge the European Research Council under the European Union’s Horizon 2020 research and innovation program (ERC-StG-Neupert-757867-PARATOP).

\bibliographystyle{apsrev4-2}
\bibliography{refs}

\begin{thebibliography}{41}%
\makeatletter
\providecommand \@ifxundefined [1]{%
 \@ifx{#1\undefined}
}%
\providecommand \@ifnum [1]{%
 \ifnum #1\expandafter \@firstoftwo
 \else \expandafter \@secondoftwo
 \fi
}%
\providecommand \@ifx [1]{%
 \ifx #1\expandafter \@firstoftwo
 \else \expandafter \@secondoftwo
 \fi
}%
\providecommand \natexlab [1]{#1}%
\providecommand \enquote  [1]{``#1''}%
\providecommand \bibnamefont  [1]{#1}%
\providecommand \bibfnamefont [1]{#1}%
\providecommand \citenamefont [1]{#1}%
\providecommand \href@noop [0]{\@secondoftwo}%
\providecommand \href [0]{\begingroup \@sanitize@url \@href}%
\providecommand \@href[1]{\@@startlink{#1}\@@href}%
\providecommand \@@href[1]{\endgroup#1\@@endlink}%
\providecommand \@sanitize@url [0]{\catcode `\\12\catcode `\$12\catcode
  `\&12\catcode `\#12\catcode `\^12\catcode `\_12\catcode `\%12\relax}%
\providecommand \@@startlink[1]{}%
\providecommand \@@endlink[0]{}%
\providecommand \url  [0]{\begingroup\@sanitize@url \@url }%
\providecommand \@url [1]{\endgroup\@href {#1}{\urlprefix }}%
\providecommand \urlprefix  [0]{URL }%
\providecommand \Eprint [0]{\href }%
\providecommand \doibase [0]{https://doi.org/}%
\providecommand \selectlanguage [0]{\@gobble}%
\providecommand \bibinfo  [0]{\@secondoftwo}%
\providecommand \bibfield  [0]{\@secondoftwo}%
\providecommand \translation [1]{[#1]}%
\providecommand \BibitemOpen [0]{}%
\providecommand \bibitemStop [0]{}%
\providecommand \bibitemNoStop [0]{.\EOS\space}%
\providecommand \EOS [0]{\spacefactor3000\relax}%
\providecommand \BibitemShut  [1]{\csname bibitem#1\endcsname}%
\let\auto@bib@innerbib\@empty
\bibitem [{\citenamefont {Troyer}\ and\ \citenamefont
  {Wiese}(2005)}]{PhysRevLett.94.170201}%
  \BibitemOpen
  \bibfield  {author} {\bibinfo {author} {\bibfnamefont {M.}~\bibnamefont
  {Troyer}}\ and\ \bibinfo {author} {\bibfnamefont {U.-J.}\ \bibnamefont
  {Wiese}},\ }\href {https://doi.org/10.1103/PhysRevLett.94.170201} {\bibfield
  {journal} {\bibinfo  {journal} {Phys. Rev. Lett.}\ }\textbf {\bibinfo
  {volume} {94}},\ \bibinfo {pages} {170201} (\bibinfo {year}
  {2005})}\BibitemShut {NoStop}%
\bibitem [{\citenamefont {Verstraete}\ \emph {et~al.}(2008)\citenamefont
  {Verstraete}, \citenamefont {Murg},\ and\ \citenamefont
  {Cirac}}]{doi:10.1080/14789940801912366}%
  \BibitemOpen
  \bibfield  {author} {\bibinfo {author} {\bibfnamefont {F.}~\bibnamefont
  {Verstraete}}, \bibinfo {author} {\bibfnamefont {V.}~\bibnamefont {Murg}},\
  and\ \bibinfo {author} {\bibfnamefont {J.}~\bibnamefont {Cirac}},\ }\href
  {https://doi.org/10.1080/14789940801912366} {\bibfield  {journal} {\bibinfo
  {journal} {Advances in Physics}\ }\textbf {\bibinfo {volume} {57}},\ \bibinfo
  {pages} {143} (\bibinfo {year} {2008})}\BibitemShut {NoStop}%
\bibitem [{\citenamefont {Carleo}\ and\ \citenamefont
  {Troyer}(2017)}]{Carleo602}%
  \BibitemOpen
  \bibfield  {author} {\bibinfo {author} {\bibfnamefont {G.}~\bibnamefont
  {Carleo}}\ and\ \bibinfo {author} {\bibfnamefont {M.}~\bibnamefont
  {Troyer}},\ }\href {https://doi.org/10.1126/science.aag2302} {\bibfield
  {journal} {\bibinfo  {journal} {Science}\ }\textbf {\bibinfo {volume}
  {355}},\ \bibinfo {pages} {602} (\bibinfo {year} {2017})}\BibitemShut
  {NoStop}%
\bibitem [{\citenamefont {Choo}\ \emph {et~al.}(2019)\citenamefont {Choo},
  \citenamefont {Neupert},\ and\ \citenamefont {Carleo}}]{PhysRevB.100.125124}%
  \BibitemOpen
  \bibfield  {author} {\bibinfo {author} {\bibfnamefont {K.}~\bibnamefont
  {Choo}}, \bibinfo {author} {\bibfnamefont {T.}~\bibnamefont {Neupert}},\ and\
  \bibinfo {author} {\bibfnamefont {G.}~\bibnamefont {Carleo}},\ }\href
  {https://doi.org/10.1103/PhysRevB.100.125124} {\bibfield  {journal} {\bibinfo
   {journal} {Phys. Rev. B}\ }\textbf {\bibinfo {volume} {100}},\ \bibinfo
  {pages} {125124} (\bibinfo {year} {2019})}\BibitemShut {NoStop}%
\bibitem [{\citenamefont {Peruzzo}\ \emph {et~al.}(2014)\citenamefont
  {Peruzzo}, \citenamefont {McClean}, \citenamefont {Shadbolt}, \citenamefont
  {Yung}, \citenamefont {Zhou}, \citenamefont {Love}, \citenamefont
  {Aspuru-Guzik},\ and\ \citenamefont {O’Brien}}]{Peruzzo_2014}%
  \BibitemOpen
  \bibfield  {author} {\bibinfo {author} {\bibfnamefont {A.}~\bibnamefont
  {Peruzzo}}, \bibinfo {author} {\bibfnamefont {J.}~\bibnamefont {McClean}},
  \bibinfo {author} {\bibfnamefont {P.}~\bibnamefont {Shadbolt}}, \bibinfo
  {author} {\bibfnamefont {M.-H.}\ \bibnamefont {Yung}}, \bibinfo {author}
  {\bibfnamefont {X.-Q.}\ \bibnamefont {Zhou}}, \bibinfo {author}
  {\bibfnamefont {P.~J.}\ \bibnamefont {Love}}, \bibinfo {author}
  {\bibfnamefont {A.}~\bibnamefont {Aspuru-Guzik}},\ and\ \bibinfo {author}
  {\bibfnamefont {J.~L.}\ \bibnamefont {O’Brien}},\ }\bibfield  {journal}
  {\bibinfo  {journal} {Nature Communications}\ }\textbf {\bibinfo {volume}
  {5}},\ \href {https://doi.org/10.1038/ncomms5213} {10.1038/ncomms5213}
  (\bibinfo {year} {2014})\BibitemShut {NoStop}%
\bibitem [{\citenamefont {McClean}\ \emph {et~al.}(2016)\citenamefont
  {McClean}, \citenamefont {Romero}, \citenamefont {Babbush},\ and\
  \citenamefont {Aspuru-Guzik}}]{McClean_2016}%
  \BibitemOpen
  \bibfield  {author} {\bibinfo {author} {\bibfnamefont {J.~R.}\ \bibnamefont
  {McClean}}, \bibinfo {author} {\bibfnamefont {J.}~\bibnamefont {Romero}},
  \bibinfo {author} {\bibfnamefont {R.}~\bibnamefont {Babbush}},\ and\ \bibinfo
  {author} {\bibfnamefont {A.}~\bibnamefont {Aspuru-Guzik}},\ }\href
  {https://doi.org/10.1088/1367-2630/18/2/023023} {\bibfield  {journal}
  {\bibinfo  {journal} {New Journal of Physics}\ }\textbf {\bibinfo {volume}
  {18}},\ \bibinfo {pages} {023023} (\bibinfo {year} {2016})}\BibitemShut
  {NoStop}%
\bibitem [{\citenamefont {Qi}\ and\ \citenamefont
  {Ranard}(2019)}]{Qi2019determininglocal}%
  \BibitemOpen
  \bibfield  {author} {\bibinfo {author} {\bibfnamefont {X.-L.}\ \bibnamefont
  {Qi}}\ and\ \bibinfo {author} {\bibfnamefont {D.}~\bibnamefont {Ranard}},\
  }\href {https://doi.org/10.22331/q-2019-07-08-159} {\bibfield  {journal}
  {\bibinfo  {journal} {{Quantum}}\ }\textbf {\bibinfo {volume} {3}},\ \bibinfo
  {pages} {159} (\bibinfo {year} {2019})}\BibitemShut {NoStop}%
\bibitem [{\citenamefont {Chertkov}\ and\ \citenamefont
  {Clark}(2018)}]{PhysRevX.8.031029}%
  \BibitemOpen
  \bibfield  {author} {\bibinfo {author} {\bibfnamefont {E.}~\bibnamefont
  {Chertkov}}\ and\ \bibinfo {author} {\bibfnamefont {B.~K.}\ \bibnamefont
  {Clark}},\ }\href {https://doi.org/10.1103/PhysRevX.8.031029} {\bibfield
  {journal} {\bibinfo  {journal} {Phys. Rev. X}\ }\textbf {\bibinfo {volume}
  {8}},\ \bibinfo {pages} {031029} (\bibinfo {year} {2018})}\BibitemShut
  {NoStop}%
\bibitem [{\citenamefont {Bairey}\ \emph {et~al.}(2019)\citenamefont {Bairey},
  \citenamefont {Arad},\ and\ \citenamefont
  {Lindner}}]{PhysRevLett.122.020504}%
  \BibitemOpen
  \bibfield  {author} {\bibinfo {author} {\bibfnamefont {E.}~\bibnamefont
  {Bairey}}, \bibinfo {author} {\bibfnamefont {I.}~\bibnamefont {Arad}},\ and\
  \bibinfo {author} {\bibfnamefont {N.~H.}\ \bibnamefont {Lindner}},\ }\href
  {https://doi.org/10.1103/PhysRevLett.122.020504} {\bibfield  {journal}
  {\bibinfo  {journal} {Phys. Rev. Lett.}\ }\textbf {\bibinfo {volume} {122}},\
  \bibinfo {pages} {020504} (\bibinfo {year} {2019})}\BibitemShut {NoStop}%
\bibitem [{\citenamefont {Valenti}\ \emph {et~al.}(2019)\citenamefont
  {Valenti}, \citenamefont {van Nieuwenburg}, \citenamefont {Huber},\ and\
  \citenamefont {Greplova}}]{Valenti_2019}%
  \BibitemOpen
  \bibfield  {author} {\bibinfo {author} {\bibfnamefont {A.}~\bibnamefont
  {Valenti}}, \bibinfo {author} {\bibfnamefont {E.}~\bibnamefont {van
  Nieuwenburg}}, \bibinfo {author} {\bibfnamefont {S.}~\bibnamefont {Huber}},\
  and\ \bibinfo {author} {\bibfnamefont {E.}~\bibnamefont {Greplova}},\
  }\bibfield  {journal} {\bibinfo  {journal} {Physical Review Research}\
  }\textbf {\bibinfo {volume} {1}},\ \href
  {https://doi.org/10.1103/physrevresearch.1.033092}
  {10.1103/physrevresearch.1.033092} (\bibinfo {year} {2019})\BibitemShut
  {NoStop}%
\bibitem [{\citenamefont {Pakrouski}(2020)}]{Pakrouski2020automaticdesignof}%
  \BibitemOpen
  \bibfield  {author} {\bibinfo {author} {\bibfnamefont {K.}~\bibnamefont
  {Pakrouski}},\ }\href {https://doi.org/10.22331/q-2020-09-02-315} {\bibfield
  {journal} {\bibinfo  {journal} {{Quantum}}\ }\textbf {\bibinfo {volume}
  {4}},\ \bibinfo {pages} {315} (\bibinfo {year} {2020})}\BibitemShut {NoStop}%
\bibitem [{\citenamefont {Dupont}\ and\ \citenamefont
  {Laflorencie}(2019)}]{PhysRevB.99.020202}%
  \BibitemOpen
  \bibfield  {author} {\bibinfo {author} {\bibfnamefont {M.}~\bibnamefont
  {Dupont}}\ and\ \bibinfo {author} {\bibfnamefont {N.}~\bibnamefont
  {Laflorencie}},\ }\href {https://doi.org/10.1103/PhysRevB.99.020202}
  {\bibfield  {journal} {\bibinfo  {journal} {Phys. Rev. B}\ }\textbf {\bibinfo
  {volume} {99}},\ \bibinfo {pages} {020202} (\bibinfo {year}
  {2019})}\BibitemShut {NoStop}%
\bibitem [{\citenamefont {Dagotto}\ and\ \citenamefont
  {Moreo}(1989)}]{PhysRevLett.63.2148}%
  \BibitemOpen
  \bibfield  {author} {\bibinfo {author} {\bibfnamefont {E.}~\bibnamefont
  {Dagotto}}\ and\ \bibinfo {author} {\bibfnamefont {A.}~\bibnamefont
  {Moreo}},\ }\href {https://doi.org/10.1103/PhysRevLett.63.2148} {\bibfield
  {journal} {\bibinfo  {journal} {Phys. Rev. Lett.}\ }\textbf {\bibinfo
  {volume} {63}},\ \bibinfo {pages} {2148} (\bibinfo {year}
  {1989})}\BibitemShut {NoStop}%
\bibitem [{\citenamefont {Schulz}\ and\ \citenamefont
  {Ziman}(1992)}]{Schulz_1992}%
  \BibitemOpen
  \bibfield  {author} {\bibinfo {author} {\bibfnamefont {H.~J.}\ \bibnamefont
  {Schulz}}\ and\ \bibinfo {author} {\bibfnamefont {T.~A.~L.}\ \bibnamefont
  {Ziman}},\ }\href {https://doi.org/10.1209/0295-5075/18/4/013} {\bibfield
  {journal} {\bibinfo  {journal} {Europhysics Letters ({EPL})}\ }\textbf
  {\bibinfo {volume} {18}},\ \bibinfo {pages} {355} (\bibinfo {year}
  {1992})}\BibitemShut {NoStop}%
\bibitem [{\citenamefont {Hu}\ \emph {et~al.}(2013)\citenamefont {Hu},
  \citenamefont {Becca}, \citenamefont {Parola},\ and\ \citenamefont
  {Sorella}}]{WJHu2013}%
  \BibitemOpen
  \bibfield  {author} {\bibinfo {author} {\bibfnamefont {W.-J.}\ \bibnamefont
  {Hu}}, \bibinfo {author} {\bibfnamefont {F.}~\bibnamefont {Becca}}, \bibinfo
  {author} {\bibfnamefont {A.}~\bibnamefont {Parola}},\ and\ \bibinfo {author}
  {\bibfnamefont {S.}~\bibnamefont {Sorella}},\ }\href
  {https://doi.org/10.1103/PhysRevB.88.060402} {\bibfield  {journal} {\bibinfo
  {journal} {Phys. Rev. B}\ }\textbf {\bibinfo {volume} {88}},\ \bibinfo
  {pages} {060402} (\bibinfo {year} {2013})}\BibitemShut {NoStop}%
\bibitem [{\citenamefont {Gong}\ \emph {et~al.}(2014)\citenamefont {Gong},
  \citenamefont {Zhu}, \citenamefont {Sheng}, \citenamefont {Motrunich},\ and\
  \citenamefont {Fisher}}]{SSGong2014}%
  \BibitemOpen
  \bibfield  {author} {\bibinfo {author} {\bibfnamefont {S.-S.}\ \bibnamefont
  {Gong}}, \bibinfo {author} {\bibfnamefont {W.}~\bibnamefont {Zhu}}, \bibinfo
  {author} {\bibfnamefont {D.~N.}\ \bibnamefont {Sheng}}, \bibinfo {author}
  {\bibfnamefont {O.~I.}\ \bibnamefont {Motrunich}},\ and\ \bibinfo {author}
  {\bibfnamefont {M.~P.~A.}\ \bibnamefont {Fisher}},\ }\href
  {https://doi.org/10.1103/PhysRevLett.113.027201} {\bibfield  {journal}
  {\bibinfo  {journal} {Phys. Rev. Lett.}\ }\textbf {\bibinfo {volume} {113}},\
  \bibinfo {pages} {027201} (\bibinfo {year} {2014})}\BibitemShut {NoStop}%
\bibitem [{\citenamefont {Jiang}\ \emph {et~al.}(2012)\citenamefont {Jiang},
  \citenamefont {Yao},\ and\ \citenamefont {Balents}}]{HCJiang2012}%
  \BibitemOpen
  \bibfield  {author} {\bibinfo {author} {\bibfnamefont {H.-C.}\ \bibnamefont
  {Jiang}}, \bibinfo {author} {\bibfnamefont {H.}~\bibnamefont {Yao}},\ and\
  \bibinfo {author} {\bibfnamefont {L.}~\bibnamefont {Balents}},\ }\href
  {https://doi.org/10.1103/PhysRevB.86.024424} {\bibfield  {journal} {\bibinfo
  {journal} {Phys. Rev. B}\ }\textbf {\bibinfo {volume} {86}},\ \bibinfo
  {pages} {024424} (\bibinfo {year} {2012})}\BibitemShut {NoStop}%
\bibitem [{\citenamefont {Sachdev}\ and\ \citenamefont
  {Bhatt}(1990)}]{Sachdev1990}%
  \BibitemOpen
  \bibfield  {author} {\bibinfo {author} {\bibfnamefont {S.}~\bibnamefont
  {Sachdev}}\ and\ \bibinfo {author} {\bibfnamefont {R.~N.}\ \bibnamefont
  {Bhatt}},\ }\href {https://doi.org/10.1103/PhysRevB.41.9323} {\bibfield
  {journal} {\bibinfo  {journal} {Phys. Rev. B}\ }\textbf {\bibinfo {volume}
  {41}},\ \bibinfo {pages} {9323} (\bibinfo {year} {1990})}\BibitemShut
  {NoStop}%
\bibitem [{\citenamefont {Mambrini}\ \emph {et~al.}(2006)\citenamefont
  {Mambrini}, \citenamefont {L\"auchli}, \citenamefont {Poilblanc},\ and\
  \citenamefont {Mila}}]{Mambrini2006}%
  \BibitemOpen
  \bibfield  {author} {\bibinfo {author} {\bibfnamefont {M.}~\bibnamefont
  {Mambrini}}, \bibinfo {author} {\bibfnamefont {A.}~\bibnamefont {L\"auchli}},
  \bibinfo {author} {\bibfnamefont {D.}~\bibnamefont {Poilblanc}},\ and\
  \bibinfo {author} {\bibfnamefont {F.}~\bibnamefont {Mila}},\ }\href
  {https://doi.org/10.1103/PhysRevB.74.144422} {\bibfield  {journal} {\bibinfo
  {journal} {Phys. Rev. B}\ }\textbf {\bibinfo {volume} {74}},\ \bibinfo
  {pages} {144422} (\bibinfo {year} {2006})}\BibitemShut {NoStop}%
\bibitem [{\citenamefont {Yu}\ and\ \citenamefont {Kao}(2012)}]{JFYu2012}%
  \BibitemOpen
  \bibfield  {author} {\bibinfo {author} {\bibfnamefont {J.-F.}\ \bibnamefont
  {Yu}}\ and\ \bibinfo {author} {\bibfnamefont {Y.-J.}\ \bibnamefont {Kao}},\
  }\href {https://doi.org/10.1103/PhysRevB.85.094407} {\bibfield  {journal}
  {\bibinfo  {journal} {Phys. Rev. B}\ }\textbf {\bibinfo {volume} {85}},\
  \bibinfo {pages} {094407} (\bibinfo {year} {2012})}\BibitemShut {NoStop}%
\bibitem [{\citenamefont {Wang}\ and\ \citenamefont
  {Sandvik}(2018)}]{LWang2018}%
  \BibitemOpen
  \bibfield  {author} {\bibinfo {author} {\bibfnamefont {L.}~\bibnamefont
  {Wang}}\ and\ \bibinfo {author} {\bibfnamefont {A.~W.}\ \bibnamefont
  {Sandvik}},\ }\href {https://doi.org/10.1103/PhysRevLett.121.107202}
  {\bibfield  {journal} {\bibinfo  {journal} {Phys. Rev. Lett.}\ }\textbf
  {\bibinfo {volume} {121}},\ \bibinfo {pages} {107202} (\bibinfo {year}
  {2018})}\BibitemShut {NoStop}%
\bibitem [{\citenamefont {Ferrari}\ and\ \citenamefont
  {Becca}(2020)}]{FFerrari2020}%
  \BibitemOpen
  \bibfield  {author} {\bibinfo {author} {\bibfnamefont {F.}~\bibnamefont
  {Ferrari}}\ and\ \bibinfo {author} {\bibfnamefont {F.}~\bibnamefont
  {Becca}},\ }\href@noop {} {\bibfield  {journal} {\bibinfo  {journal} {arXiv
  preprint arXiv:2005.12941}\ } (\bibinfo {year} {2020})}\BibitemShut {NoStop}%
\bibitem [{\citenamefont {Nomura}\ and\ \citenamefont
  {Imada}(2020{\natexlab{a}})}]{YNomura2020}%
  \BibitemOpen
  \bibfield  {author} {\bibinfo {author} {\bibfnamefont {Y.}~\bibnamefont
  {Nomura}}\ and\ \bibinfo {author} {\bibfnamefont {M.}~\bibnamefont {Imada}},\
  }\href@noop {} {\bibfield  {journal} {\bibinfo  {journal} {arXiv preprint
  arXiv:2005.14142}\ } (\bibinfo {year} {2020}{\natexlab{a}})}\BibitemShut
  {NoStop}%
\bibitem [{\citenamefont {Cybenko}(1989)}]{cybenko1989}%
  \BibitemOpen
  \bibfield  {author} {\bibinfo {author} {\bibfnamefont {G.}~\bibnamefont
  {Cybenko}},\ }\href {https://doi.org/10.1007/BF02551274} {\bibfield
  {journal} {\bibinfo  {journal} {Mathematics of Control, Signals and Systems}\
  }\textbf {\bibinfo {volume} {2}},\ \bibinfo {pages} {303} (\bibinfo {year}
  {1989})}\BibitemShut {NoStop}%
\bibitem [{\citenamefont {Bishop}(1995)}]{bishop-book}%
  \BibitemOpen
  \bibfield  {author} {\bibinfo {author} {\bibfnamefont {C.}~\bibnamefont
  {Bishop}},\ }\href@noop {} {\emph {\bibinfo {title} {Neural Networks for
  Pattern Recognition}}}\ (\bibinfo  {publisher} {Oxford University Press},\
  \bibinfo {year} {1995})\BibitemShut {NoStop}%
\bibitem [{\citenamefont {Deng}\ \emph {et~al.}(2017)\citenamefont {Deng},
  \citenamefont {Li},\ and\ \citenamefont {Das~Sarma}}]{deng2017prx}%
  \BibitemOpen
  \bibfield  {author} {\bibinfo {author} {\bibfnamefont {D.-L.}\ \bibnamefont
  {Deng}}, \bibinfo {author} {\bibfnamefont {X.}~\bibnamefont {Li}},\ and\
  \bibinfo {author} {\bibfnamefont {S.}~\bibnamefont {Das~Sarma}},\ }\href
  {https://doi.org/10.1103/PhysRevX.7.021021} {\bibfield  {journal} {\bibinfo
  {journal} {Phys. Rev. X}\ }\textbf {\bibinfo {volume} {7}},\ \bibinfo {pages}
  {021021} (\bibinfo {year} {2017})}\BibitemShut {NoStop}%
\bibitem [{\citenamefont {Nomura}\ \emph {et~al.}(2017)\citenamefont {Nomura},
  \citenamefont {Darmawan}, \citenamefont {Yamaji},\ and\ \citenamefont
  {Imada}}]{PhysRevB.96.205152}%
  \BibitemOpen
  \bibfield  {author} {\bibinfo {author} {\bibfnamefont {Y.}~\bibnamefont
  {Nomura}}, \bibinfo {author} {\bibfnamefont {A.~S.}\ \bibnamefont
  {Darmawan}}, \bibinfo {author} {\bibfnamefont {Y.}~\bibnamefont {Yamaji}},\
  and\ \bibinfo {author} {\bibfnamefont {M.}~\bibnamefont {Imada}},\ }\href
  {https://doi.org/10.1103/PhysRevB.96.205152} {\bibfield  {journal} {\bibinfo
  {journal} {Phys. Rev. B}\ }\textbf {\bibinfo {volume} {96}},\ \bibinfo
  {pages} {205152} (\bibinfo {year} {2017})}\BibitemShut {NoStop}%
\bibitem [{\citenamefont {Carleo}\ \emph {et~al.}(2018)\citenamefont {Carleo},
  \citenamefont {Nomura},\ and\ \citenamefont
  {Imada}}]{carleo_nomura_imada_2018}%
  \BibitemOpen
  \bibfield  {author} {\bibinfo {author} {\bibfnamefont {G.}~\bibnamefont
  {Carleo}}, \bibinfo {author} {\bibfnamefont {Y.}~\bibnamefont {Nomura}},\
  and\ \bibinfo {author} {\bibfnamefont {M.}~\bibnamefont {Imada}},\ }\bibfield
   {journal} {\bibinfo  {journal} {Nature Communications}\ }\textbf {\bibinfo
  {volume} {9}},\ \href {https://doi.org/10.1038/s41467-018-07520-3}
  {10.1038/s41467-018-07520-3} (\bibinfo {year} {2018})\BibitemShut {NoStop}%
\bibitem [{\citenamefont {Chen}\ \emph {et~al.}(2018)\citenamefont {Chen},
  \citenamefont {Cheng}, \citenamefont {Xie}, \citenamefont {Wang},\ and\
  \citenamefont {Xiang}}]{PhysRevB.97.085104}%
  \BibitemOpen
  \bibfield  {author} {\bibinfo {author} {\bibfnamefont {J.}~\bibnamefont
  {Chen}}, \bibinfo {author} {\bibfnamefont {S.}~\bibnamefont {Cheng}},
  \bibinfo {author} {\bibfnamefont {H.}~\bibnamefont {Xie}}, \bibinfo {author}
  {\bibfnamefont {L.}~\bibnamefont {Wang}},\ and\ \bibinfo {author}
  {\bibfnamefont {T.}~\bibnamefont {Xiang}},\ }\href
  {https://doi.org/10.1103/PhysRevB.97.085104} {\bibfield  {journal} {\bibinfo
  {journal} {Phys. Rev. B}\ }\textbf {\bibinfo {volume} {97}},\ \bibinfo
  {pages} {085104} (\bibinfo {year} {2018})}\BibitemShut {NoStop}%
\bibitem [{\citenamefont {Glasser}\ \emph {et~al.}(2018)\citenamefont
  {Glasser}, \citenamefont {Pancotti}, \citenamefont {August}, \citenamefont
  {Rodriguez},\ and\ \citenamefont {Cirac}}]{PhysRevX.8.011006}%
  \BibitemOpen
  \bibfield  {author} {\bibinfo {author} {\bibfnamefont {I.}~\bibnamefont
  {Glasser}}, \bibinfo {author} {\bibfnamefont {N.}~\bibnamefont {Pancotti}},
  \bibinfo {author} {\bibfnamefont {M.}~\bibnamefont {August}}, \bibinfo
  {author} {\bibfnamefont {I.~D.}\ \bibnamefont {Rodriguez}},\ and\ \bibinfo
  {author} {\bibfnamefont {J.~I.}\ \bibnamefont {Cirac}},\ }\href
  {https://doi.org/10.1103/PhysRevX.8.011006} {\bibfield  {journal} {\bibinfo
  {journal} {Phys. Rev. X}\ }\textbf {\bibinfo {volume} {8}},\ \bibinfo {pages}
  {011006} (\bibinfo {year} {2018})}\BibitemShut {NoStop}%
\bibitem [{\citenamefont {Saito}\ and\ \citenamefont
  {Kato}(2018)}]{doi:10.7566/JPSJ.87.014001}%
  \BibitemOpen
  \bibfield  {author} {\bibinfo {author} {\bibfnamefont {H.}~\bibnamefont
  {Saito}}\ and\ \bibinfo {author} {\bibfnamefont {M.}~\bibnamefont {Kato}},\
  }\href {https://doi.org/10.7566/JPSJ.87.014001} {\bibfield  {journal}
  {\bibinfo  {journal} {Journal of the Physical Society of Japan}\ }\textbf
  {\bibinfo {volume} {87}},\ \bibinfo {pages} {014001} (\bibinfo {year}
  {2018})}\BibitemShut {NoStop}%
\bibitem [{\citenamefont {Choo}\ \emph {et~al.}(2018)\citenamefont {Choo},
  \citenamefont {Carleo}, \citenamefont {Regnault},\ and\ \citenamefont
  {Neupert}}]{PhysRevLett.121.167204}%
  \BibitemOpen
  \bibfield  {author} {\bibinfo {author} {\bibfnamefont {K.}~\bibnamefont
  {Choo}}, \bibinfo {author} {\bibfnamefont {G.}~\bibnamefont {Carleo}},
  \bibinfo {author} {\bibfnamefont {N.}~\bibnamefont {Regnault}},\ and\
  \bibinfo {author} {\bibfnamefont {T.}~\bibnamefont {Neupert}},\ }\href
  {https://doi.org/10.1103/PhysRevLett.121.167204} {\bibfield  {journal}
  {\bibinfo  {journal} {Phys. Rev. Lett.}\ }\textbf {\bibinfo {volume} {121}},\
  \bibinfo {pages} {167204} (\bibinfo {year} {2018})}\BibitemShut {NoStop}%
\bibitem [{\citenamefont {Kochkov}\ and\ \citenamefont
  {Clark}(2018)}]{kochkov2018variational}%
  \BibitemOpen
  \bibfield  {author} {\bibinfo {author} {\bibfnamefont {D.}~\bibnamefont
  {Kochkov}}\ and\ \bibinfo {author} {\bibfnamefont {B.~K.}\ \bibnamefont
  {Clark}},\ }\href@noop {} {\bibinfo {title} {Variational optimization in the
  ai era: Computational graph states and supervised wave-function
  optimization}} (\bibinfo {year} {2018}),\ \Eprint
  {https://arxiv.org/abs/1811.12423} {arXiv:1811.12423 [cond-mat.str-el]}
  \BibitemShut {NoStop}%
\bibitem [{\citenamefont {Luo}\ and\ \citenamefont {Clark}(2019)}]{Luo_2019}%
  \BibitemOpen
  \bibfield  {author} {\bibinfo {author} {\bibfnamefont {D.}~\bibnamefont
  {Luo}}\ and\ \bibinfo {author} {\bibfnamefont {B.~K.}\ \bibnamefont
  {Clark}},\ }\bibfield  {journal} {\bibinfo  {journal} {Physical Review
  Letters}\ }\textbf {\bibinfo {volume} {122}},\ \href
  {https://doi.org/10.1103/physrevlett.122.226401}
  {10.1103/physrevlett.122.226401} (\bibinfo {year} {2019})\BibitemShut
  {NoStop}%
\bibitem [{\citenamefont {Pastori}\ \emph {et~al.}(2019)\citenamefont
  {Pastori}, \citenamefont {Kaubruegger},\ and\ \citenamefont
  {Budich}}]{PhysRevB.99.165123}%
  \BibitemOpen
  \bibfield  {author} {\bibinfo {author} {\bibfnamefont {L.}~\bibnamefont
  {Pastori}}, \bibinfo {author} {\bibfnamefont {R.}~\bibnamefont
  {Kaubruegger}},\ and\ \bibinfo {author} {\bibfnamefont {J.~C.}\ \bibnamefont
  {Budich}},\ }\href {https://doi.org/10.1103/PhysRevB.99.165123} {\bibfield
  {journal} {\bibinfo  {journal} {Phys. Rev. B}\ }\textbf {\bibinfo {volume}
  {99}},\ \bibinfo {pages} {165123} (\bibinfo {year} {2019})}\BibitemShut
  {NoStop}%
\bibitem [{\citenamefont {Sharir}\ \emph {et~al.}(2020)\citenamefont {Sharir},
  \citenamefont {Levine}, \citenamefont {Wies}, \citenamefont {Carleo},\ and\
  \citenamefont {Shashua}}]{sharir2020deep}%
  \BibitemOpen
  \bibfield  {author} {\bibinfo {author} {\bibfnamefont {O.}~\bibnamefont
  {Sharir}}, \bibinfo {author} {\bibfnamefont {Y.}~\bibnamefont {Levine}},
  \bibinfo {author} {\bibfnamefont {N.}~\bibnamefont {Wies}}, \bibinfo {author}
  {\bibfnamefont {G.}~\bibnamefont {Carleo}},\ and\ \bibinfo {author}
  {\bibfnamefont {A.}~\bibnamefont {Shashua}},\ }\href@noop {} {\bibfield
  {journal} {\bibinfo  {journal} {Physical Review Letters}\ }\textbf {\bibinfo
  {volume} {124}},\ \bibinfo {pages} {020503} (\bibinfo {year}
  {2020})}\BibitemShut {NoStop}%
\bibitem [{\citenamefont {Nomura}\ and\ \citenamefont
  {Imada}(2020{\natexlab{b}})}]{nomura2020diractype}%
  \BibitemOpen
  \bibfield  {author} {\bibinfo {author} {\bibfnamefont {Y.}~\bibnamefont
  {Nomura}}\ and\ \bibinfo {author} {\bibfnamefont {M.}~\bibnamefont {Imada}},\
  }\href@noop {} {\bibinfo {title} {Dirac-type nodal spin liquid revealed by
  machine learning}} (\bibinfo {year} {2020}{\natexlab{b}}),\ \Eprint
  {https://arxiv.org/abs/2005.14142} {arXiv:2005.14142 [cond-mat.str-el]}
  \BibitemShut {NoStop}%
\bibitem [{\citenamefont {Gao}\ and\ \citenamefont {Duan}(2017)}]{Gao_2017}%
  \BibitemOpen
  \bibfield  {author} {\bibinfo {author} {\bibfnamefont {X.}~\bibnamefont
  {Gao}}\ and\ \bibinfo {author} {\bibfnamefont {L.-M.}\ \bibnamefont {Duan}},\
  }\bibfield  {journal} {\bibinfo  {journal} {Nature Communications}\ }\textbf
  {\bibinfo {volume} {8}},\ \href {https://doi.org/10.1038/s41467-017-00705-2}
  {10.1038/s41467-017-00705-2} (\bibinfo {year} {2017})\BibitemShut {NoStop}%
\bibitem [{\citenamefont {Levine}\ \emph {et~al.}(2019)\citenamefont {Levine},
  \citenamefont {Sharir}, \citenamefont {Cohen},\ and\ \citenamefont
  {Shashua}}]{yoav2019prl}%
  \BibitemOpen
  \bibfield  {author} {\bibinfo {author} {\bibfnamefont {Y.}~\bibnamefont
  {Levine}}, \bibinfo {author} {\bibfnamefont {O.}~\bibnamefont {Sharir}},
  \bibinfo {author} {\bibfnamefont {N.}~\bibnamefont {Cohen}},\ and\ \bibinfo
  {author} {\bibfnamefont {A.}~\bibnamefont {Shashua}},\ }\href
  {https://doi.org/10.1103/PhysRevLett.122.065301} {\bibfield  {journal}
  {\bibinfo  {journal} {Phys. Rev. Lett.}\ }\textbf {\bibinfo {volume} {122}},\
  \bibinfo {pages} {065301} (\bibinfo {year} {2019})}\BibitemShut {NoStop}%
\bibitem [{\citenamefont {Carleo}\ \emph {et~al.}(2019)\citenamefont {Carleo},
  \citenamefont {Choo}, \citenamefont {Hofmann}, \citenamefont {Smith},
  \citenamefont {Westerhout}, \citenamefont {Alet}, \citenamefont {Davis},
  \citenamefont {Efthymiou}, \citenamefont {Glasser}, \citenamefont {Lin},
  \citenamefont {Mauri}, \citenamefont {Mazzola}, \citenamefont {Mendl},
  \citenamefont {{van Nieuwenburg}}, \citenamefont {O’Reilly}, \citenamefont
  {Théveniaut}, \citenamefont {Torlai}, \citenamefont {Vicentini},\ and\
  \citenamefont {Wietek}}]{netket}%
  \BibitemOpen
  \bibfield  {author} {\bibinfo {author} {\bibfnamefont {G.}~\bibnamefont
  {Carleo}}, \bibinfo {author} {\bibfnamefont {K.}~\bibnamefont {Choo}},
  \bibinfo {author} {\bibfnamefont {D.}~\bibnamefont {Hofmann}}, \bibinfo
  {author} {\bibfnamefont {J.~E.}\ \bibnamefont {Smith}}, \bibinfo {author}
  {\bibfnamefont {T.}~\bibnamefont {Westerhout}}, \bibinfo {author}
  {\bibfnamefont {F.}~\bibnamefont {Alet}}, \bibinfo {author} {\bibfnamefont
  {E.~J.}\ \bibnamefont {Davis}}, \bibinfo {author} {\bibfnamefont
  {S.}~\bibnamefont {Efthymiou}}, \bibinfo {author} {\bibfnamefont
  {I.}~\bibnamefont {Glasser}}, \bibinfo {author} {\bibfnamefont {S.-H.}\
  \bibnamefont {Lin}}, \bibinfo {author} {\bibfnamefont {M.}~\bibnamefont
  {Mauri}}, \bibinfo {author} {\bibfnamefont {G.}~\bibnamefont {Mazzola}},
  \bibinfo {author} {\bibfnamefont {C.~B.}\ \bibnamefont {Mendl}}, \bibinfo
  {author} {\bibfnamefont {E.}~\bibnamefont {{van Nieuwenburg}}}, \bibinfo
  {author} {\bibfnamefont {O.}~\bibnamefont {O’Reilly}}, \bibinfo {author}
  {\bibfnamefont {H.}~\bibnamefont {Théveniaut}}, \bibinfo {author}
  {\bibfnamefont {G.}~\bibnamefont {Torlai}}, \bibinfo {author} {\bibfnamefont
  {F.}~\bibnamefont {Vicentini}},\ and\ \bibinfo {author} {\bibfnamefont
  {A.}~\bibnamefont {Wietek}},\ }\href
  {https://doi.org/https://doi.org/10.1016/j.softx.2019.100311} {\bibfield
  {journal} {\bibinfo  {journal} {SoftwareX}\ }\textbf {\bibinfo {volume}
  {10}},\ \bibinfo {pages} {100311} (\bibinfo {year} {2019})}\BibitemShut
  {NoStop}%
\bibitem [{\citenamefont {Vieijra}\ \emph {et~al.}(2020)\citenamefont
  {Vieijra}, \citenamefont {Casert}, \citenamefont {Nys}, \citenamefont
  {De~Neve}, \citenamefont {Haegeman}, \citenamefont {Ryckebusch},\ and\
  \citenamefont {Verstraete}}]{PhysRevLett.124.097201}%
  \BibitemOpen
  \bibfield  {author} {\bibinfo {author} {\bibfnamefont {T.}~\bibnamefont
  {Vieijra}}, \bibinfo {author} {\bibfnamefont {C.}~\bibnamefont {Casert}},
  \bibinfo {author} {\bibfnamefont {J.}~\bibnamefont {Nys}}, \bibinfo {author}
  {\bibfnamefont {W.}~\bibnamefont {De~Neve}}, \bibinfo {author} {\bibfnamefont
  {J.}~\bibnamefont {Haegeman}}, \bibinfo {author} {\bibfnamefont
  {J.}~\bibnamefont {Ryckebusch}},\ and\ \bibinfo {author} {\bibfnamefont
  {F.}~\bibnamefont {Verstraete}},\ }\href
  {https://doi.org/10.1103/PhysRevLett.124.097201} {\bibfield  {journal}
  {\bibinfo  {journal} {Phys. Rev. Lett.}\ }\textbf {\bibinfo {volume} {124}},\
  \bibinfo {pages} {097201} (\bibinfo {year} {2020})}\BibitemShut {NoStop}%
\end{thebibliography}%


\begin{thebibliography}{41}%
\makeatletter
\providecommand \@ifxundefined [1]{%
 \@ifx{#1\undefined}
}%
\providecommand \@ifnum [1]{%
 \ifnum #1\expandafter \@firstoftwo
 \else \expandafter \@secondoftwo
 \fi
}%
\providecommand \@ifx [1]{%
 \ifx #1\expandafter \@firstoftwo
 \else \expandafter \@secondoftwo
 \fi
}%
\providecommand \natexlab [1]{#1}%
\providecommand \enquote  [1]{``#1''}%
\providecommand \bibnamefont  [1]{#1}%
\providecommand \bibfnamefont [1]{#1}%
\providecommand \citenamefont [1]{#1}%
\providecommand \href@noop [0]{\@secondoftwo}%
\providecommand \href [0]{\begingroup \@sanitize@url \@href}%
\providecommand \@href[1]{\@@startlink{#1}\@@href}%
\providecommand \@@href[1]{\endgroup#1\@@endlink}%
\providecommand \@sanitize@url [0]{\catcode `\\12\catcode `\$12\catcode
  `\&12\catcode `\#12\catcode `\^12\catcode `\_12\catcode `\%12\relax}%
\providecommand \@@startlink[1]{}%
\providecommand \@@endlink[0]{}%
\providecommand \url  [0]{\begingroup\@sanitize@url \@url }%
\providecommand \@url [1]{\endgroup\@href {#1}{\urlprefix }}%
\providecommand \urlprefix  [0]{URL }%
\providecommand \Eprint [0]{\href }%
\providecommand \doibase [0]{https://doi.org/}%
\providecommand \selectlanguage [0]{\@gobble}%
\providecommand \bibinfo  [0]{\@secondoftwo}%
\providecommand \bibfield  [0]{\@secondoftwo}%
\providecommand \translation [1]{[#1]}%
\providecommand \BibitemOpen [0]{}%
\providecommand \bibitemStop [0]{}%
\providecommand \bibitemNoStop [0]{.\EOS\space}%
\providecommand \EOS [0]{\spacefactor3000\relax}%
\providecommand \BibitemShut  [1]{\csname bibitem#1\endcsname}%
\let\auto@bib@innerbib\@empty
\bibitem [{\citenamefont {Choo}\ \emph {et~al.}(2019)\citenamefont {Choo},
  \citenamefont {Neupert},\ and\ \citenamefont {Carleo}}]{PhysRevB.100.125124}%
  \BibitemOpen
  \bibfield  {author} {\bibinfo {author} {\bibfnamefont {K.}~\bibnamefont
  {Choo}}, \bibinfo {author} {\bibfnamefont {T.}~\bibnamefont {Neupert}},\ and\
  \bibinfo {author} {\bibfnamefont {G.}~\bibnamefont {Carleo}},\ }\href
  {https://doi.org/10.1103/PhysRevB.100.125124} {\bibfield  {journal} {\bibinfo
   {journal} {Phys. Rev. B}\ }\textbf {\bibinfo {volume} {100}},\ \bibinfo
  {pages} {125124} (\bibinfo {year} {2019})}\BibitemShut {NoStop}%

\end{thebibliography}%

\end{document}


\title{Supplemental Material - Hamiltonian reconstruction as metric for variational studies}
\author{Kevin Zhang}
\affiliation{Laboratory of Atomic and Solid State Physics, Cornell University, 142 Sciences Drive, Ithaca NY 14853-2501, USA}
\author{Samuel Lederer}
\affiliation{Laboratory of Atomic and Solid State Physics, Cornell University, 142 Sciences Drive, Ithaca NY 14853-2501, USA}
\author{Kenny Choo}
\affiliation{Department of Physics, University of Zurich, Winterthurerstrasse 190, 8057 Zurich, Switzerland}
\author{Titus Neupert}
\affiliation{Department of Physics, University of Zurich, Winterthurerstrasse 190, 8057 Zurich, Switzerland}
\author{Giuseppe Carleo}
\affiliation{Computational Quantum Science Laboratory, \'{e}cole polytechnique f\'{e}d\'{e}rale de Lausanne, Route Cantonale, 1015 Lausanne, Switzerland}
\author{Eun-Ah Kim}
\affiliation{Laboratory of Atomic and Solid State Physics, Cornell University, 142 Sciences Drive, Ithaca NY 14853-2501, USA}
\date{January 2021}

\maketitle

\section{Neural network architectures and training}
\label{sec:wf}

In this section, we will explain how we constructed and trained our variational ansatze.
First, we describe a basic neural network to motivate the more complex architectures that we used.
Then, we explicitly construct the convolutional neural network (CNN) and restricted Boltzmann machine (RBM) architectures that we used in our study.
Finally, we explain how we trained and symmetrized the wavefunctions.

One of the simplest examples of a neural network that can be used to parametrize a variational wavefunction is a single-layer perceptron.
Although we did not use this architecture, it serves as an illustrative example of neural network wavefunctions in general.
The single-layer perceptron consists of one transformation (known as a dense layer)
\begin{equation}
    w_i(\{ v_{i'} \}) = g\left(\sum_j \bm{W}_{ij} v_j + b_i\right),
\end{equation}
where $W$ is a complex matrix of tunable parameters (``weights"), $\vec{b}$ is a complex vector of tunable parameters (``biases"), and $g$ represents a non-linear firing function.
The wavefunction corresponding to this single-layer network is
\begin{equation}
    \bra{\{\sigma_i\}}\ket{\Psi} = \sum_i w_i(\{ \sigma_{i'} \})
\end{equation}
Generically, a multi-layer perceptron can be constructed by composing several of these ``linear" layers.
Examples of common algorithms used to optimize these constructions are stochastic gradient descent or Adam.
In our training, we used stochastic gradient descent with a learning rate of $0.001$.

\subsection{Convolutional neural network}
In a convolutional layer, both the input and output have an additional ``channel" dimension on top of the spatial indices.
Thus, each stage of intermediate values is a three-dimensional tensor, with one channel index and two spatial indices.
The input values (spin configuration) are interpreted as just consisting of a single channel.
A convolutional layer from $N$ channels to $M$ channels is given by
\begin{equation}
    w_{m,(i,j)}(\{v_{n,i',j'}\}) = g\left(\sum_{n}^N \sum_{x=0}^{L-1} \sum_{y=0}^{L-1} K_{n,(x,y)} v_{n,(i+x,j+y)} + b_m\right),
\end{equation}
where the indices of the output vector $w$ are $m$ for the channel, and $(i, j)$ for the spatial position.
The input vector $v$ is similarly indexed.
As with the dense layer, $g(x) = \ln{\cosh{x}}$ is the nonlinearity.
The $K$ parameters are called the ``kernel'' of the transformation, and there is one kernel for each output channel.
Each kernel is a $L$ by $L$ by $N$ tensor of weights, which attempts to capture some spatially local feature of the input values.
$L$ is the size of the kernel.
The construction of the convolutional layer is explicitly translationally invariant, which respects the symmetry of the periodic J1-J2 model we are solving.
The full convolutional network is given by the composition of convolutional layers,
\begin{equation}
    \bra{\sigma_{x, y}}\ket{\Psi} = \sum_{w,i,j} w_{m,(i,j)}\left(\{v_{n,(i',j')}(...(\sigma_{x,y}))\}\right)
\end{equation}
The outputs of the final layer were summed to obtain the output of the network.
The network we used consists of three convolutional layers, with channel counts 6, 4, 2, and kernel sizes 4, 4, 2, respectively, amounting to a total of 236 parameters.

\subsection{Restricted Boltzmann machine}
The single-layer restricted Boltzmann machine is characterized by the wavefunction
\begin{equation}
    \bra{\{ \sigma_i \}}\ket{\Psi} = e^{\sum_i a_i \sigma_i} \prod_j^M \cosh \left(\sum_i W_{ij} \sigma_i + b_j \right)
\end{equation}
where $M$ represents the number of hidden nodes.
We imposed the spatial symmetries of translation and rotation on the parameters $a_i$, $W_{ij}$, $b_j$, reducing the total number of free parameters by a factor of 16 (the size of the system).
With a hidden node count of 160, this wavefunction has a total of 171 parameters.

\subsection{Sign rotation and symmetrization}

\begin{figure*}
\centering
    \subfigure[] {
    \centering
    \includegraphics[width=.3\textwidth]{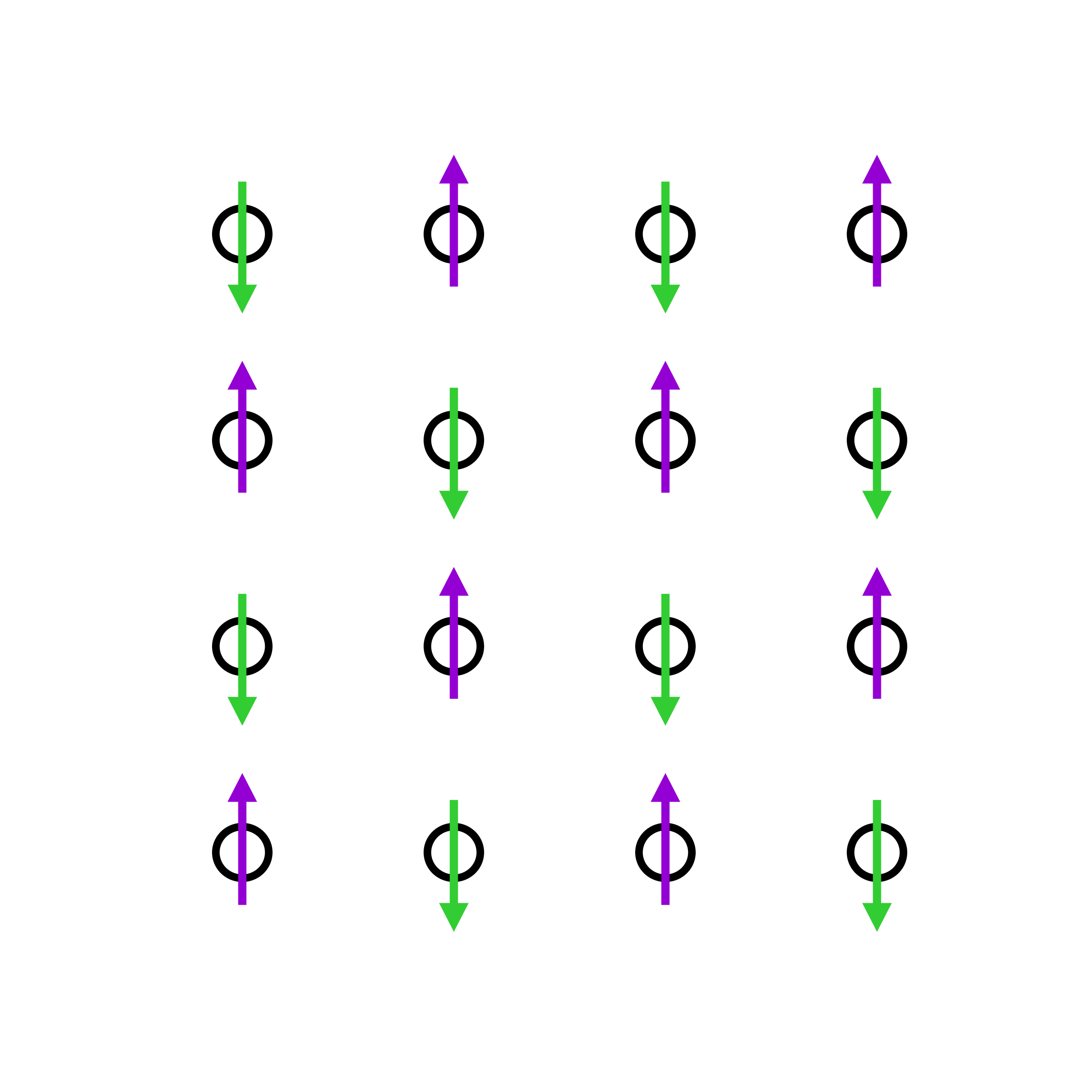}
    }
    \subfigure[] {
    \centering
    \includegraphics[width=.3\textwidth]{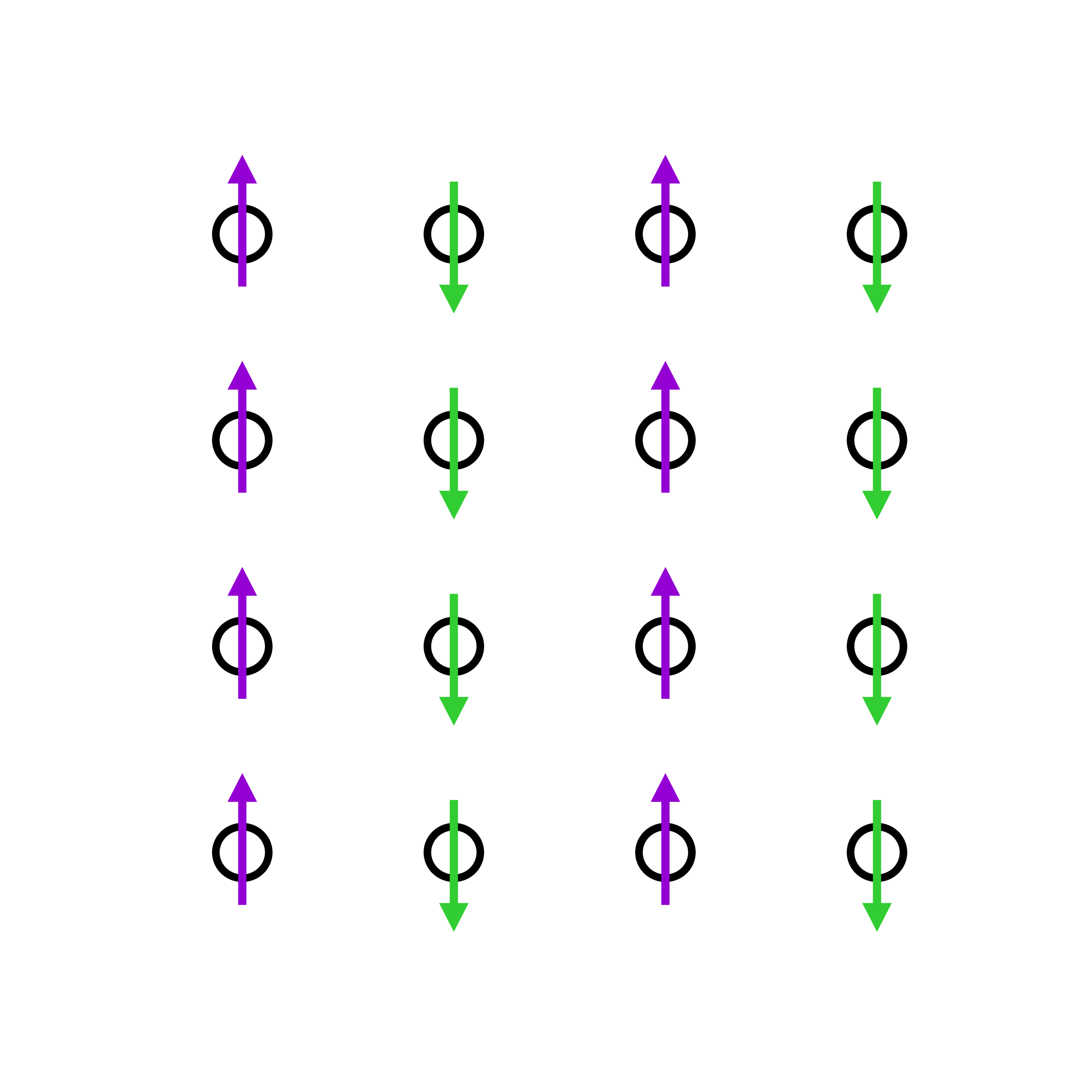}
    }
    \caption{Diagrams of the classical spin structure, used for in-plane wavefunction rotation, for a) N\'{e}el and b) stripe order.}
    \label{fig:spin}
\end{figure*}

Since the successful optimization of wavefunctions depends on a sign structure being pre-imposed, the wavefunctions were rotated in-plane:
\begin{equation}
     \ket{\Psi} \to \left( \bigotimes_{i \in A} e^{-i \pi \sigma^i_z / 2} \right) \ket{\Psi},
\end{equation}
where $A$ denotes a subset of lattice sites corresponding to either N\'{e}el or stripe order (\autoref{fig:spin}).
The N\'{e}el rotation was used for $J_2 \leq 0.5$ and the striped rotation used otherwise.

Finally, to ensure that the final wavefunctions respect point group symmetries and time reversal, we used the same procedure as in \cite{PhysRevB.100.125124}.
For an Abelian symmetry $C$, with an irrep $\Gamma$, the (unnormalized) symmetrization of the wavefunction $\ket{\Psi}$ under $\Gamma$ is given by
\begin{equation}
    \ket{\Psi_S} = \sum_{r=0}^{|C| - 1} \omega^r c^r \ket{\Psi},
\end{equation}
where $\omega$ is the character of $\Gamma$ and $c$ is the generator of $\Gamma$.
For the $C_4$ rotation group, there are four possible irreps with characters $\pm 1$ and $\pm i$; for time reversal symmetry, there are two irreps with characters $\pm 1$.
We chose the combination of representations that yielded the lowest ground state energy, and in all cases, the energy was lowest with the irrep with character 1 for both $C_4$ and $T$.

\section{Hamiltonian reconstruction}
\subsection{Benchmarks}
With the exact ground state of the Heisenberg model ($J_2=0$) obtained via exact diagonalization, Hamiltonian reconstruction indeed yields the correct Hamiltonian in all three subspaces we considered down to machine precision, as shown in Table~I.

\begin{table}[ht]
\centering
\begin{tabular}{ |c||c| }

 \hline
 $\mathcal{H}[\mathcal{O}]$ & Reconstructed parameter \\
 \hline
 $H[\delta J_2]$  & $\delta J_2 = -6.5 \times 10^{-15} $ \\
 $H[J_3]$  & $J_3 = -3.18 \times 10^{-15} $ \\
 $H[\alpha]$ & $\alpha = 3.73 \times 10^{-15} $ \\
 \hline
\end{tabular}

\caption{Benchmark reconstructions for wavefunctions obtained via exact diagonalization for the $J_1$-$J_2$ model in the limit $J_2 = 0$.
In all cases, the reconstruction yields the original $J_1$-$J_2$ Hamiltonian to within machine precision.}
\label{tab:table}
\end{table}

\subsection{Multi-dimensional reconstructions}

In our studies, we chose to restrict the target Hamiltonian spaces to two-dimensional spaces.
Attempting to perform reconstruction into large Hamiltonian spaces led to unidentifiable features.
Here, we summarize the results of such a high-dimensional reconstruction, in the space given by

\begin{equation}
     H[\alpha, \beta, \delta J_2] = H_{J_1J_2} + \alpha \sum_{\left< i, j\right>} S^z_i S^z_{j} + \beta \sum_{\llangle i, j\rrangle} S^z_i S^z_{j} + \delta J_2 \sum_{\llangle i, j\rrangle} \vec{S}_i \cdot \vec{S}_j,
\end{equation}

\begin{figure*}[htp]
\subfigure[] {
    \centering
    \includegraphics[width=.3\textwidth]{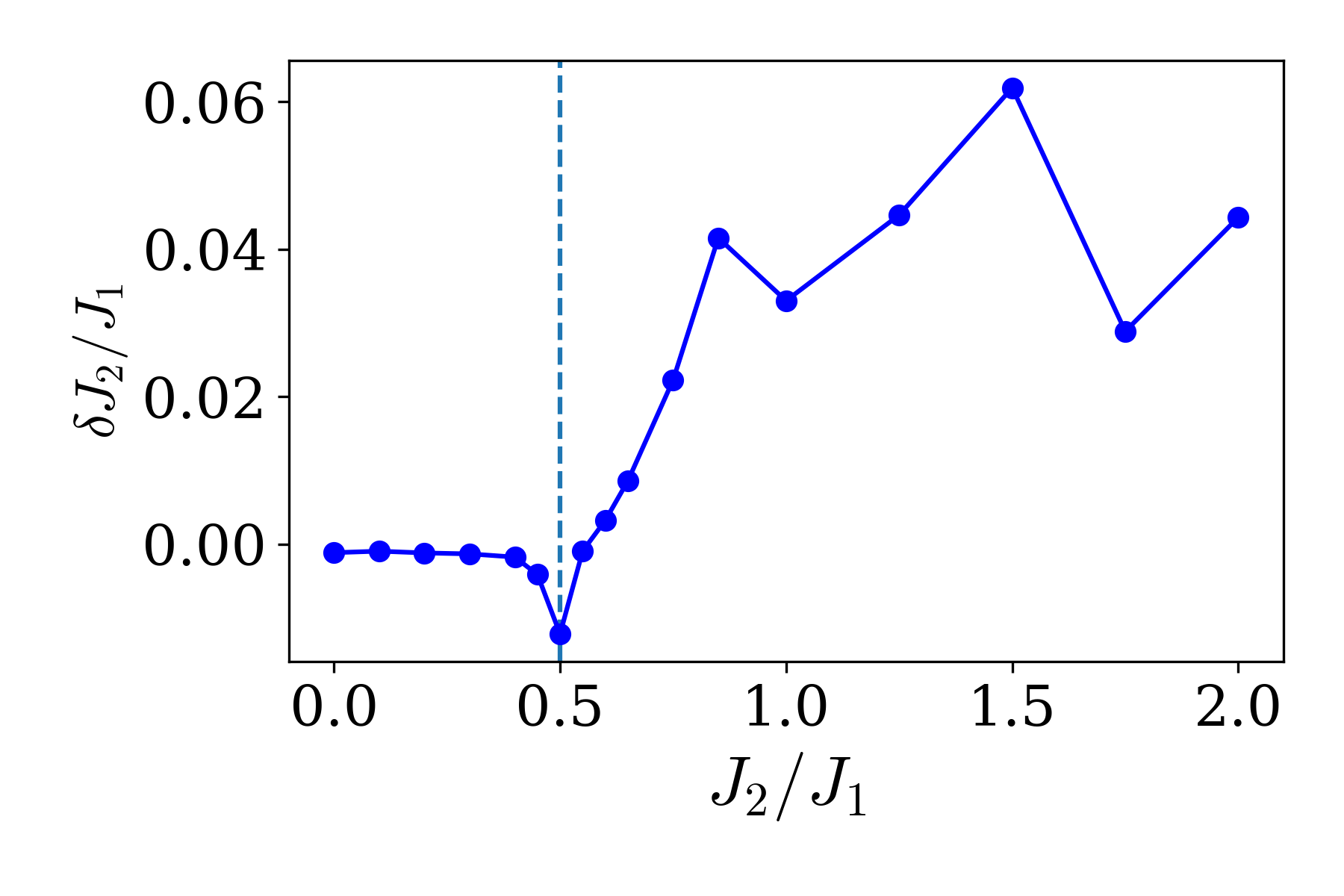}
    }\subfigure[] {
    \centering
    \includegraphics[width=.3\textwidth]{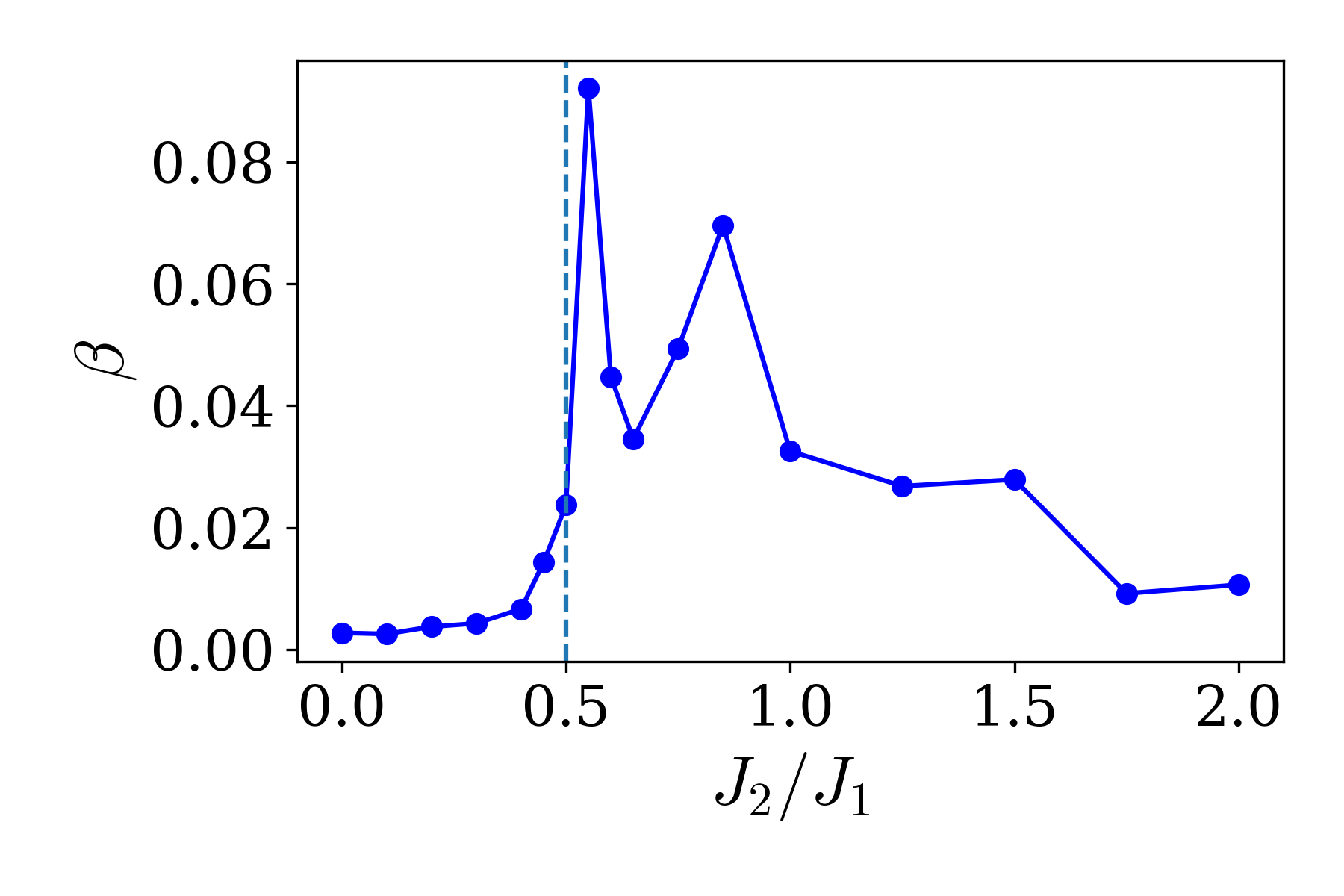}
    }\subfigure[] {
    \centering
    \includegraphics[width=.3\textwidth]{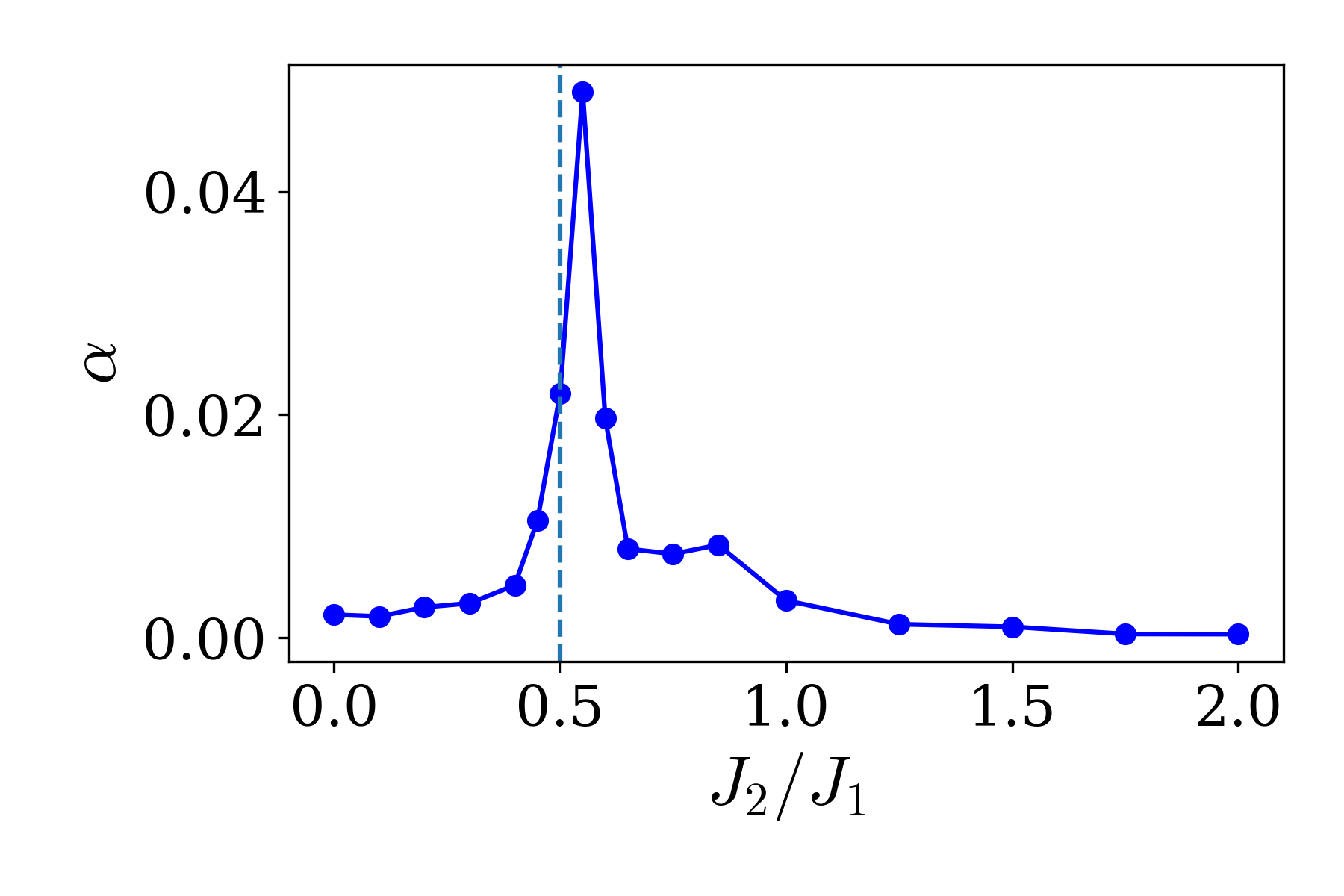}
    }
    \caption{Parameters of the reconstructed Hamiltonians into a higher-dimensional target space of CNN wavefunctions.
    a) Reconstructed nearest-neighbor coupling $\delta J_2/J_1$.
    b) The reconstructed nearest neighbor easy-axis anisotropy $\alpha/J_1$.
    c) The reconstructed next-nearest neighbor easy-axis anisotropy $\beta/J_1$.}
    \label{fig:H1}
\end{figure*}

\autoref{fig:H1} shows reconstruction results for this four-dimensional Hamiltonian space. Here, an additional unexplained peak at $J_2/J_1 = 0.75$ can be observed in the $\alpha$ parameter which is not present in reconstructions of two-dimensional target spaces.

Further, multi-dimensional target spaces including a $J_3$ parameter, e.g.
\begin{equation}
    H[\alpha, \beta, \delta J_2, J_3] = H[\alpha, \beta, \delta J_2] + J_3 \sum_{\langle i, j\rangle_3} \vec{S}_i \cdot \vec{S}_j,
\end{equation}
result in degeneracy of the correlation matrix's lowest eigenvalues. In these cases, there is no unique best Hamiltonian making interpretation of the reconstruction difficult.
Thus, for clarity, we restricted our analysis to two-dimensional target spaces.

\subsection{Monte Carlo sampling}

Various methods can be used when evaluating the elements of the quantum covariance matrix.
Here, we present results of Hamiltonian reconstruction using Monte Carlo sampling for correlation functions.
The model in question is the same $4 \times 4$ square lattice $J_1$-$J_2$ Heisenberg model, with $J_2/J_1 = 0.5$, i.e., the classical high frustration point.
We only present results here for the reconstruction into the space $H[\delta J_2]$ using the convolutional neural network wavefunction.

\begin{figure*}[htp]
\centering
    \subfigure[] {
    \centering
    \includegraphics[width=0.3\textwidth]{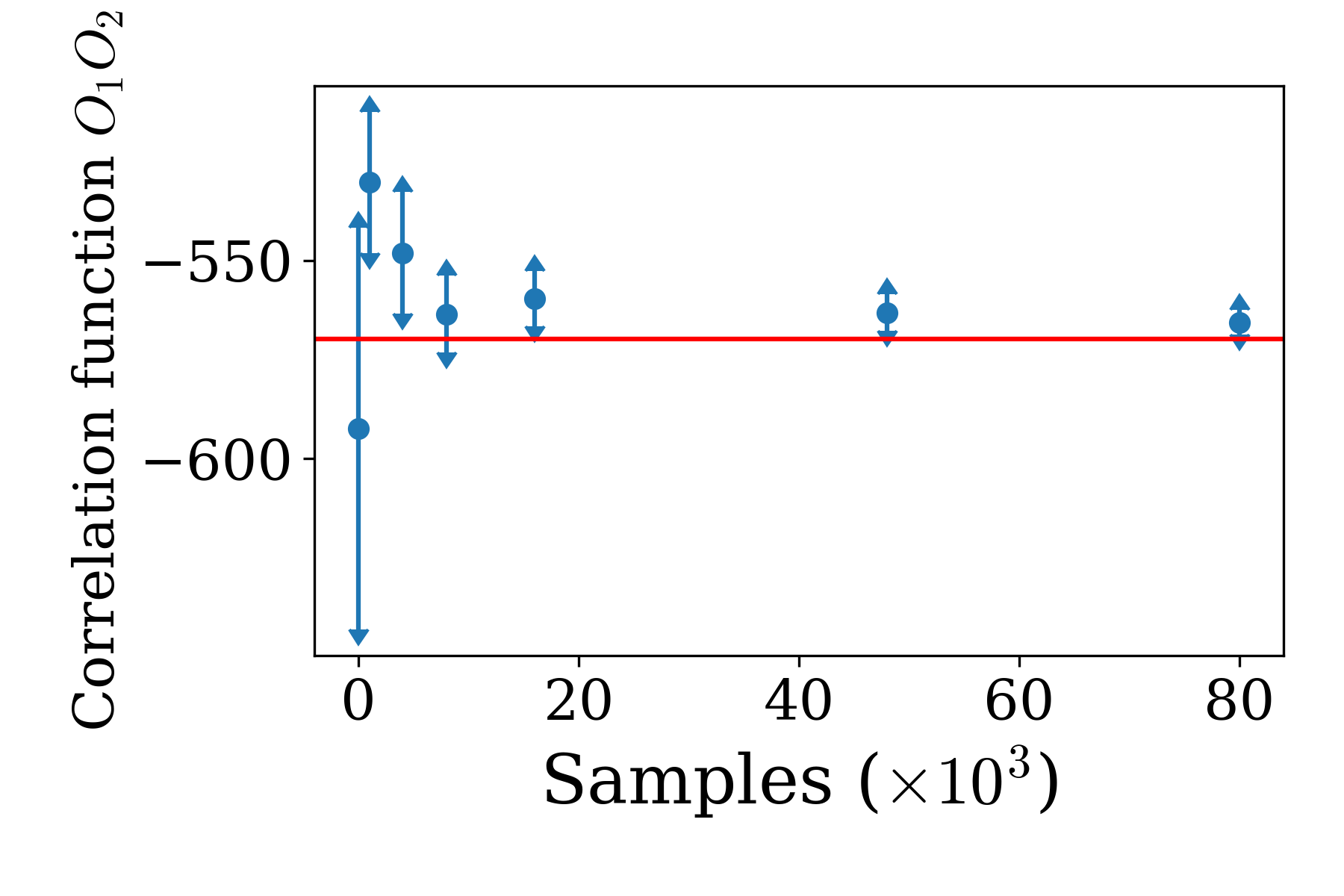}
    }\subfigure[] {
    \centering
    \includegraphics[width=0.3\textwidth]{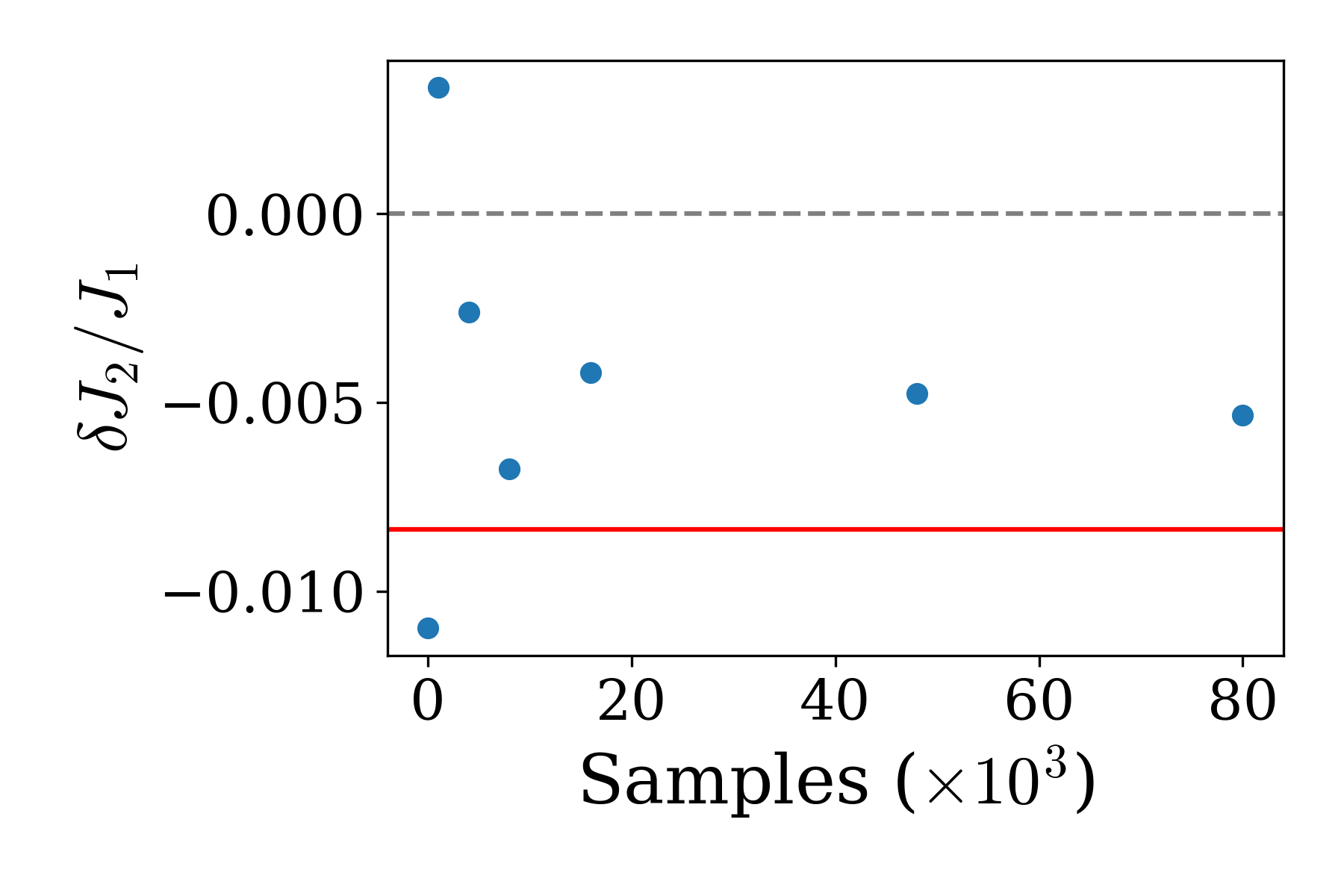}
    }\subfigure[] {
    \centering
    \includegraphics[width=0.3\textwidth]{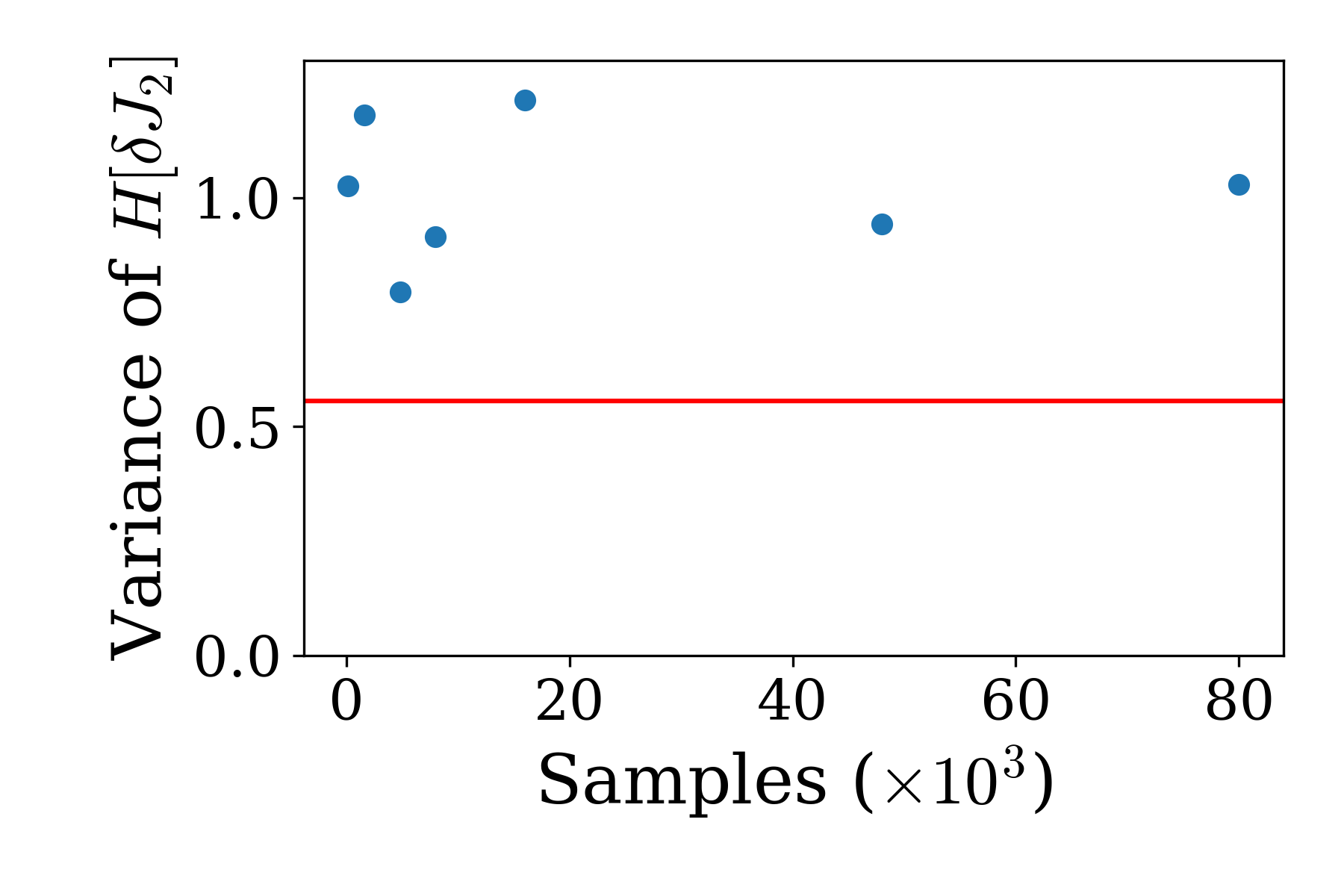}
    }
    
    \caption{Results of Hamiltonian reconstruction using Monte Carlo sampling for operator correlation functions at $J_2/J_1 = 0.5$, into the space $H[\delta J_2]$. In all plots, the horizontal red line represents the true value as obtained through direct integration of the correlation functions.
    a) Correlation function $\langle{H_{J1J2} \sum_{\llangle i, j\rrangle} \vec{S}_i \cdot \vec{S}_j \rangle}$ as measured by Monte Carlo sampling. The error bars represent the standard deviation of the estimates for the correlation function.
    b) The reconstructed $\delta J_2$ parameter, which converges slower than the previous correlation function.
    c) The variance of the variational wavefunction $\ket{\Psi}$ under the reconstructed Hamiltonian. Here, the variance does not converge within the number of Monte Carlo samples that we used.}
    \label{fig:mc}
\end{figure*}

The reconstructed $\delta J_2$ parameter showed a slower rate of convergence (\autoref{fig:mc}(b)) than that of the shown correlation function (\autoref{fig:mc}(a)) with the number of Monte Carlo samples.
Further, the variance of the input wavefunction under the reconstructed wavefunction did not converge for the range of Monte Carlo samples that we used (\autoref{fig:mc}(c)).
Considering these points, we chose for our study to restrict our analysis to correlation functions evaluated explicitly. As a result, we limited ourselves to systems small enough that they were amenable to exact diagonalization.

\section{Correlations in the $J_1$-$J_2$-$J_3$ model}

Here, we present measurements of spin-spin correlation functions on the ground states of $J_1$-$J_2$-$J_3$ models obtained via exact diagonalization.
We define the spin-spin correlation function for wavevector $q$ as:
\begin{equation}
\label{eq:corr}
    S(\vec{q}) = \frac{1}{N} \sum_{i, j} e^{i \vec{q}\cdot (\vec{r}_i - \vec{r_j})} \langle \vec{S}_i \cdot \vec{S}_j \rangle,
\end{equation}
where $N=16$ is the system size.

\begin{figure*}[htp]
    \centering
    \subfigure[] {
    \centering
    \includegraphics[width=.4\textwidth]{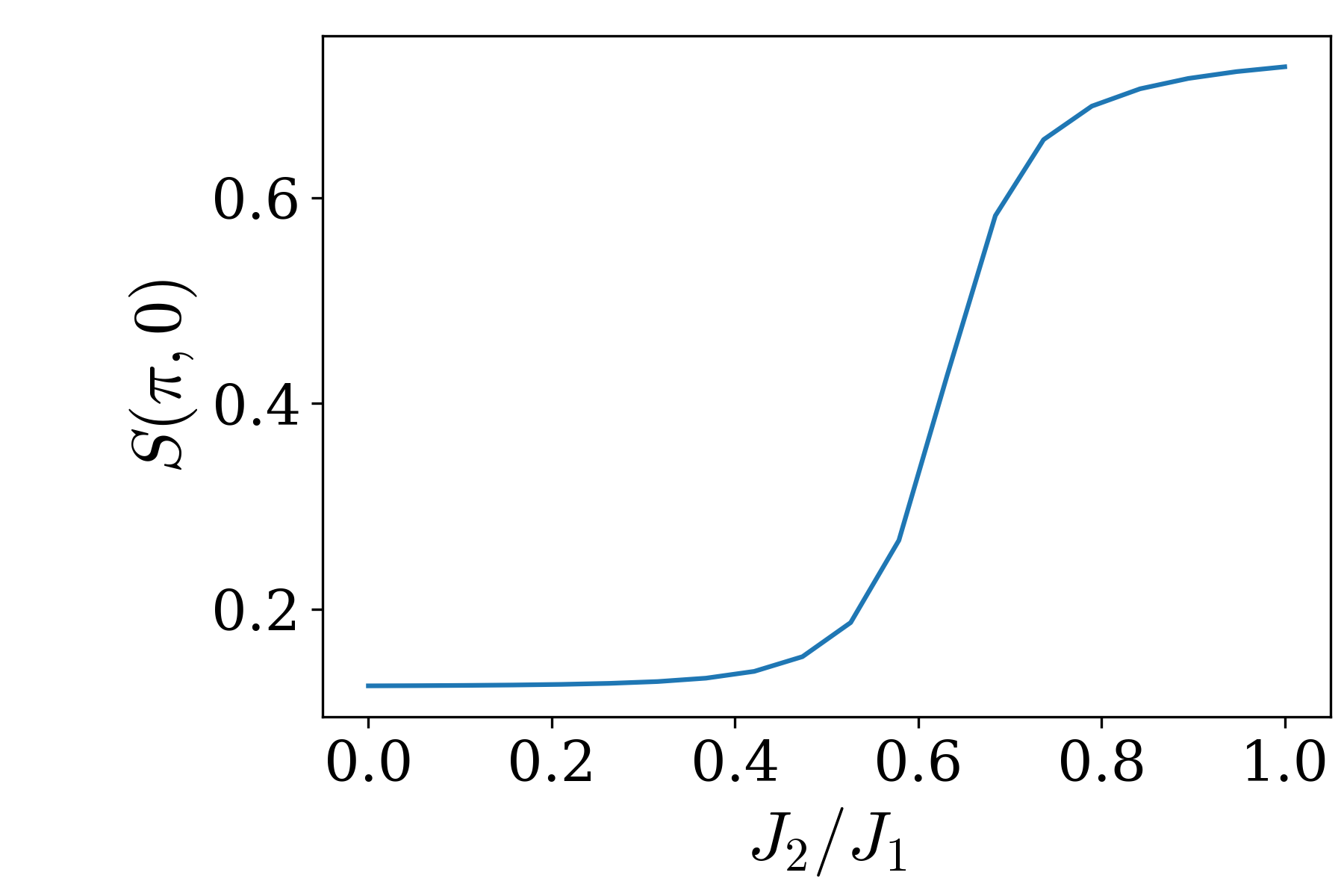}
    }
    \subfigure[] {
    \centering
    \includegraphics[width=.4\textwidth]{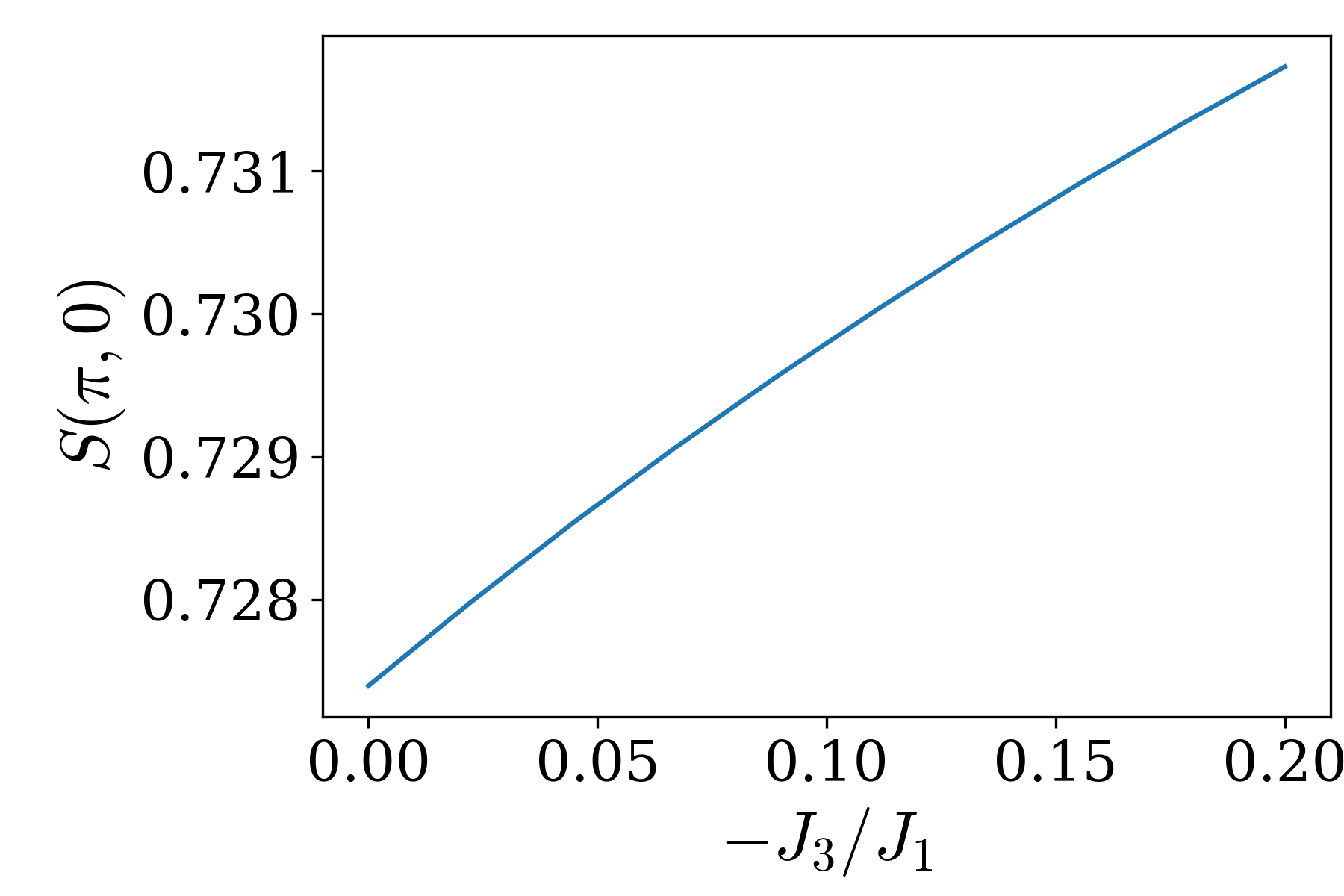}
    }
    
    \caption{Spin-spin correlation functions, as defined in \autoref{eq:corr}, of the exact ground state of $J_1$-$J_2$-$J_3$ models. a) $S(\pi, 0)$ for $J_3 = 0$, showing how the stripe order is enhanced with increasing $J_2$. b) Similarly, $S(\pi, 0)$ at $J_2/J_1 = 1$ for ferromagnetic $J_3$ is enhanced.
    }
    \label{fig:corr}
\end{figure*}

As can be seen from \autoref{fig:corr}, an increase in the $J_2$ parameter in the large $J_2$ regime leads to enhanced stripe order (\autoref{fig:corr}(a)). Also, small negative values of $J_3$ of similar magnitude to those reconstructed from our variational wavefunctions enhance the stripe order for states in the same regime (\autoref{fig:corr}(b)).
The stripe order parameter here is the spin-spin correlation function with ordering vector $(\pi, 0)$ or $(0, \pi)$.
These observations show that a ferromagnetic $J_3$ parameter, as well as positive $\delta J_2$, act to suppress quantum fluctuations and enhance the ground state order of the model in the large $J_2$ regime.

\bibliographystyle{apsrev4-2}
\bibliography{refs}